\documentclass [
 preprint,
 superscriptaddress,
 amsmath,amssymb,
 aps]
{revtex4-2}
\usepackage[margin=1.5cm]{geometry} 
\usepackage{setspace}
\usepackage{etoolbox}

\AtBeginDocument{%
  \pretocmd{\caption}{\setstretch{1}}{}{}%
}

\makeatletter
\newcommand{\beginesupplementary}{%
  \clearpage
  \let\orig@figurename\figurename
  \let\orig@thefigure\thefigure
  \let\orig@thepage\thepage
  \thispagestyle{empty}%
  \begin{center}
    {\large \textbf{Supplementary Information}\par}
  \end{center}
  \vspace{2em}%
  \setcounter{page}{1}%
  \renewcommand{\thepage}{S\arabic{page}}%
  \setcounter{figure}{0}%
  \renewcommand{\figurename}{Fig.}%
  \renewcommand{\thefigure}{S\arabic{figure}}%
}

\makeatletter
\newcommand{\beginextendeddata}{%
  \clearpage
  \let\orig@figurename\figurename
  \let\orig@thefigure\thefigure
  \let\orig@thepage\thepage
  \thispagestyle{empty}%
  \begin{center}
    {\large \textbf{Extended data figures}\par}
  \end{center}
  \vspace{2em}%
  \setcounter{page}{1}%
  \renewcommand{\thepage}{ED\arabic{page}}%
  \setcounter{figure}{0}%
  \renewcommand{\figurename}{Extended Data Fig.}%
  \renewcommand{\thefigure}{\arabic{figure}}%
}

\makeatother



\newcommand{\edf}[1]{\hyperref[#1]{Extended Data Fig.~\ref{#1}}}

\newcommand{\subedf}[2]{\hyperref[#1]{Extended Data Fig.~\ref{#1}#2}}
\newcommand{\subref}[2]{\hyperref[#1]{\ref{#1}#2}}

\usepackage [colorlinks=true,allcolors=blue]{hyperref}
\usepackage{comment}
\usepackage{graphicx}
\usepackage{dcolumn}
\usepackage{bm}
\usepackage{siunitx}
\DeclareSIUnit{\rad}{rad}
\usepackage{soul}
\usepackage{placeins}
 \DeclareSIUnit\bar{bar}
\usepackage{lineno}
\usepackage{textgreek}
\usepackage [english]{babel}

\usepackage{natbib}
\usepackage{physics}
\begin{document}
\setlength{\tabcolsep}{5pt}
\renewcommand{\arraystretch}{1.4}

\renewcommand{\abstractname}{\normalfont\scshape{}}

\title{A fluctuation-free pathway for a topological magnetic phase transition}
\author{Riccardo Battistelli}
\affiliation{Helmholtz-Zentrum Berlin für Materialien und Energie GmbH, 14109 Berlin, Germany}
\affiliation{Experimental Physics V, Center for Electronic Correlations and Magnetism, University of Augsburg, 86159 Augsburg, Germany}
\author{Lukas K\"orber}
\affiliation{Helmholtz-Zentrum Berlin für Materialien und Energie GmbH, 14109 Berlin, Germany}
\affiliation{Institute for Molecules and Materials, Radboud University, Nijmegen, 6525 AJ, Netherlands}
\affiliation{Institut f\"ur Ionenstrahlphysik und Materialforschung, Helmholtz-Zentrum Dresden - Rossendorf, 01328, Dresden, Germany}
\author{Kai Litzius}
\affiliation{Experimental Physics V, Center for Electronic Correlations and Magnetism, University of Augsburg, 86159 Augsburg, Germany}
\author{Matthieu Grelier}
\affiliation{Laboratoire Albert Fert CNRS, Thales, Université Paris-Saclay, 91767 Palaiseau, France}
\author{Krishnanjana Puzhekadavil Joy}
\affiliation{Helmholtz-Zentrum Berlin für Materialien und Energie GmbH, 14109 Berlin, Germany}
\affiliation{Experimental Physics V, Center for Electronic Correlations and Magnetism, University of Augsburg, 86159 Augsburg, Germany}
\author{Michael Schneider}
\affiliation{Max Born Institute for Nonlinear Optics and Short Pulse Spectroscopy, 12489 Berlin, Germany}
\author{Steffen Wittrock}
\affiliation{Helmholtz-Zentrum Berlin für Materialien und Energie GmbH, 14109 Berlin, Germany}
\author{Daniel Metternich}
\affiliation{Helmholtz-Zentrum Berlin für Materialien und Energie GmbH, 14109 Berlin, Germany}
\affiliation{Max Born Institute for Nonlinear Optics and Short Pulse Spectroscopy, 12489 Berlin, Germany}
\author{Tamer Karaman}
\affiliation{Experimental Physics V, Center for Electronic Correlations and Magnetism, University of Augsburg, 86159 Augsburg, Germany}
\author{Lisa-Marie Kern}
\author{Christopher Klose}
\affiliation{Max Born Institute for Nonlinear Optics and Short Pulse Spectroscopy, 12489 Berlin, Germany}
\author{Simone Finizio}
\affiliation{Swiss Light Source, Paul Scherrer Institute, 5232 Villigen PSI, Switzerland.}
\author{Josefin Fuchs}
\affiliation{Max Born Institute for Nonlinear Optics and Short Pulse Spectroscopy, 12489 Berlin, Germany}
\author{Christian M. G\"unther}
\affiliation{Technische Universit\"at Berlin, Zentraleinrichtung Elektronenmikroskopie (ZELMI), 10623 Berlin, Germany}
\author{Tim A. Butcher}
\affiliation{Max Born Institute for Nonlinear Optics and Short Pulse Spectroscopy, 12489 Berlin, Germany}
\affiliation{Swiss Light Source, Paul Scherrer Institute, 5232 Villigen PSI, Switzerland.}
\author{Karel Proke\v{s}}
\affiliation{Helmholtz-Zentrum Berlin für Materialien und Energie GmbH, 14109 Berlin, Germany}
\author{Raluca Boltje}
\affiliation{Helmholtz-Zentrum Berlin für Materialien und Energie GmbH, 14109 Berlin, Germany}
\affiliation{Experimental Physics V, Center for Electronic Correlations and Magnetism, University of Augsburg, 86159 Augsburg, Germany}
\author{Manas Patra}
\affiliation{Experimental Physics V, Center for Electronic Correlations and Magnetism, University of Augsburg, 86159 Augsburg, Germany}
\author{Sebastian Wintz}
\affiliation{Helmholtz-Zentrum Berlin für Materialien und Energie GmbH, 14109 Berlin, Germany}
\author{Markus Weigand}
\affiliation{Helmholtz-Zentrum Berlin für Materialien und Energie GmbH, 14109 Berlin, Germany}
\author{Sascha Petz}
\affiliation{Helmholtz-Zentrum Berlin für Materialien und Energie GmbH, 14109 Berlin, Germany}
\author{Horia Popescu}
\affiliation{Synchrotron SOLEIL, 91192 Saint-Aubin, France}
\author{J\"org Raabe}
\affiliation{Swiss Light Source, Paul Scherrer Institute, 5232 Villigen PSI, Switzerland.}
\author{Nicolas Jaouen}
\affiliation{Synchrotron SOLEIL, 91192 Saint-Aubin, France}
\author{Stefan Eisebitt}
\affiliation{Max Born Institute for Nonlinear Optics and Short Pulse Spectroscopy, 12489 Berlin, Germany}
\affiliation{Technische Universit\"at Berlin, Institute of Physics and Astronomy, 10623 Berlin, Germany}
\author{Vincent Cros}
\affiliation{Laboratoire Albert Fert CNRS, Thales, Université Paris-Saclay, 91767 Palaiseau, France}
\author{Bastian Pfau}
\affiliation{Max Born Institute for Nonlinear Optics and Short Pulse Spectroscopy, 12489 Berlin, Germany}
\affiliation{Technische Universit\"at Berlin, Institute of Physics and Astronomy, 10623 Berlin, Germany}
\author{Johan H. Mentink}
\affiliation{Institute for Molecules and Materials, Radboud University, Nijmegen, 6525 AJ, Netherlands}
\author{Nicolas Reyren}
\affiliation{Laboratoire Albert Fert CNRS, Thales, Université Paris-Saclay, 91767 Palaiseau, France}
\author{Felix B\"uttner}
\email{felix.buettner@uni-a.de}
\affiliation{Helmholtz-Zentrum Berlin für Materialien und Energie GmbH, 14109 Berlin, Germany}
\affiliation{Experimental Physics V, Center for Electronic Correlations and Magnetism, University of Augsburg, 86159 Augsburg, Germany}

\date{\today}
\begin{abstract}
\textbf{
Topological magnetic textures are particle-like spin configurations stabilized by competing interactions.
Their formation is commonly attributed to fluctuation-driven, first-order nucleation processes requiring activation over a topological energy barrier.
Here, we demonstrate an alternative barrier- and fluctuation-free pathway for nucleating topological magnetic textures, triggered in our experiments by an excitation-induced spin reorientation transition.
By combining x-ray imaging, scattering and micromagnetic simulations, we show that the system follows a deterministic cascade of symmetry-breaking phase transitions after excitation.
First, the system undergoes a second-order phase transition from a homogeneous state to weak stripe domains, then a first-order transition to topologically trivial bubbles, and finally a topological switching event into skyrmionic textures.
Through simulations, we generalize our findings and demonstrate that this pathway is active in a vast range of low-anisotropy materials.
This previously unrecognized, spontaneous transition pathway suggests strategies for rapid, low-energy generation of topological spin textures and points to a general role of intrinsic modulational instabilities in phase transitions beyond magnetism.
}

\end{abstract}
\maketitle 

Emergent domain patterns are commonly observed in systems with competing interactions~\cite{seul1995}.
Such structures are central to condensed matter physics because of their role in mediating phase transitions and linking microscopic interactions to observable macroscopic phenomena (for example, transport properties in Hall-effect measurements).
In magnetic thin films, these patterns can exhibit non-trivial topology and particle-like character, as in the case of merons~\cite{shinjo2000}, skyrmions~\cite{nagaosa2013,fert2017} and hopfions~\cite{kent2021}.
These topological textures attracted attention because of their rich dynamics~\cite{reichhardt2022}, their role in seeding the formation and collapse of complex magnetic domain patterns~\cite{cape1971} and potential application in spintronic technologies~\cite{luo2021}.
Yet the mechanism of their creation, especially topological switching, is still an open field of research.

Due to the change in topological charge, topological texture formation is typically considered a first-order phase transition, requiring activation over energy barriers~\cite{buttner2018,bernand-mantel2025}.
This interpretation is supported by experiments in which skyrmion nucleation relied on heterogeneity in either the excitation or material~\cite{finizio2019, fallon2020, buttner2017} or on strong thermal activation~\cite{legrand2017, wang2020, buttner2021, lemesh2018}, as expected for a first order phase transition.
However, several studies on low-anisotropy films~\cite{he2018,fallarino2019,denker2023} and under tilted fields~\cite{ denker2023,moon2021,wu2021,salikhov2022, titze2025} suggest that skyrmionic textures can form uniformly without strong thermal excitation.
For example, lattices of skyrmionic textures were nucleated in Fe$\mid$Gd multilayers by destabilizing non-chiral stripes via tilted field cycling~\cite{je2020}, or appeared spontaneously after saturation in films with layer-dependent properties such as low-anisotropy Ir$\mid$Fe$\mid$Co$\mid$Pt~\cite{yildirim2022} and Pt$\mid$Co$\mid$Al multilayers~\cite{grelier2022}.
These observations point to a deterministic pathway for topological domain formation that does not rely on fluctuations or defects, although the underlying mechanism remains unclear.

In this work, we uncover a universal pathway for controlled, barrier-free nucleation of topological textures in low-anisotropy magnetic thin films.
Experiments and simulations show that a transient reduction of the anisotropy in these materials can trigger a spin-reorientation transition into a quasi-homogeneous state with strong in-plane character, which evolves, in the presence of a symmetry-breaking out-of-plane magnetic field, into a dense lattice of skyrmion-like textures once the anisotropy is restored.
We show that this process requires neither fluctuations nor defects, challenging common understanding of topological domain nucleation~\cite{buttner2021,zhang2023}.
We identify the driving mechanism as a cascade of modulational instabilities followed by a topological switching event.
This sequence enables deterministic domain nucleation and topological switching via intrinsic energetic instabilities, providing a new framework for magnetic topological phase transitions beyond the conventional fluctuation-activated paradigm.

\begin{figure*}[h]
\includegraphics[width=180 mm]{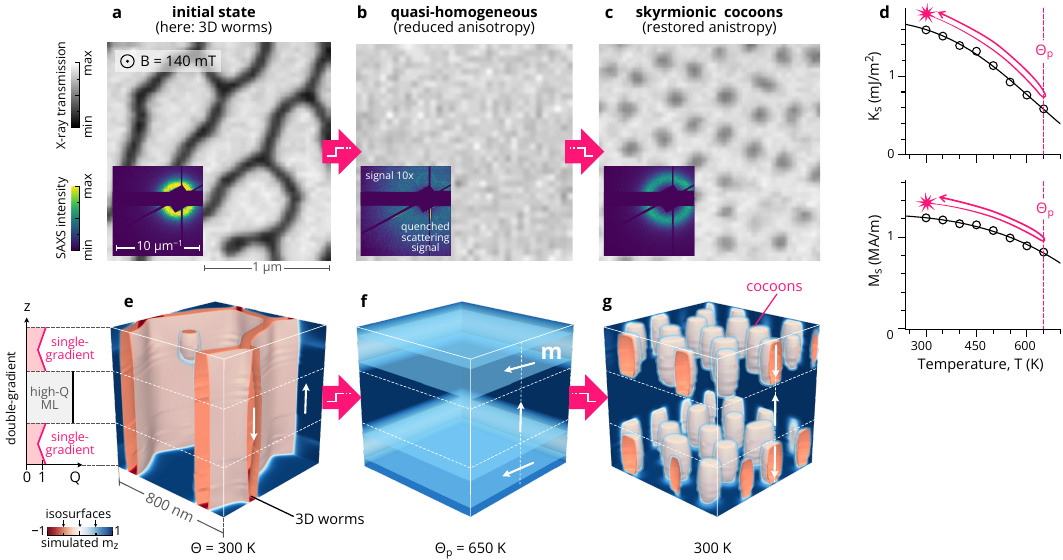}
\caption{
\textbf{Evidence of spin-reorientation-transition-mediated topological switching.}
\textbf{a}-\textbf{c,} Scanning transmission x-ray microscopy images of magnetic states in a double-gradient magnetic multilayer before (a), during (b), and after (c) quasi-static heating of the material.
Insets: SAXS patterns of the cocoon-hosting single-gradient section of the material during the same process. The SAXS signal in (b) has been multiplied by a factor 10 to improve visibility.
\textbf{d,} Temperature-dependence of the interfacial magnetic anisotropy $K_\mathrm{s}$ and saturation magnetization $M_\mathrm{s}$ of our materials.
The pink arrows are a guide to the eye, exemplifying the change of the micromagnetic parameters during a thermal excitation pulse to a transient pseudo-temperature $\Theta_\mathrm{p}$.
See \hyperref[sec:Methods]{Methods} for details.
\textbf{e}-\textbf{g,} Micromagnetic simulation of the thermal excitation process from room temperature to a maximum transient pseudo-temperature $\Theta_\mathrm{p}$, akin to the one displayed in (\textbf{a-c}).
Isosurfaces display the normalized out-of-plane magnetization $m_z={M_z}/{M_\mathrm{s}}$. Blue: $m_z=+0.5$, white: $m_z=0$, red: $m_z=-0.5$.
The plot on the side displays the layer-dependent quality factor $Q$ of a typical double-gradient magnetic multilayer at room temperature.
}
\label{fig:fig1}
\end{figure*}

\section*{Observation of the transition}

Our study focuses on [Pt$\mid$Co$\mid$Al]\textsubscript{$\times N$} aperiodic magnetic multilayers (AMLs)~\cite{grelier2022}, an ideal testbed because these materials can combine both first-order and barrier-free topological switching.
Two related architectures are used: single-gradient and double-gradient AMLs.
In single-gradient AMLs, the Co thickness changes gradually from layer to layer, producing an anisotropy gradient with quality factors $Q<1$ in the inner layers ($Q=\frac{2 K_\text{u}}{\mu_0 M_\text{s}^2}$, where $\mu_0$ is the vacuum permeability and $K_\text{u}$ and $M_\text{s}$ are the uniaxial anisotropy and saturation magnetization of a given magnetic layer).
These stacks readily form dense arrays of bubble-like skyrmionic textures upon lowering the external out-of-plane ($z$ axis) magnetic field from saturation~\cite{grelier2022}.
The vertical gradient in $Q$ confines the resulting topological textures to the film interior, giving them a prolate three-dimensional shape; these are referred to as cocoons (Fig.~\subref{fig:fig1}{c,g}).
Two single-gradient sections can sandwich a high-$Q$ multilayer to create a double-gradient architecture (see Fig.~\subref{fig:fig1}{e}).
Such composite stacks can host cocoons in the single-gradient sections along with 3D worms that penetrate the entire stack.
In x-ray transmission microscopy, 3D worms can be distinguished from cocoons by their stronger magnetic contrast (see Figs.~\subref{fig:fig1}{a,c}).
As previously shown~\cite{grelier2022}, the two textures possess distinct nucleation behaviours: cocoons nucleate spontaneously in dense lattices during field cycling, whereas 3D worms originate from a few nucleation centers and expand into extended maze-like patterns, a hallmark of first-order nucleation-and-growth switching~\cite{cape1971} (see~\edf{fig:EXTFIG1}).
AMLs therefore provide a unique platform to study distinct nucleation mechanisms under identical external conditions.

In our experiments, we reliably produced dense cocoon phases in such AMLs not only via field cycling but also via thermal excitation across all timescales, from femtosecond laser pulses over nanosecond current pulses to continuous Joule-heating currents (see~\edf{fig:Fig_transitions_means2}).
The latter offered us the opportunity to characterize the high-temperature transient state with quasi-static scattering, microscopy and magnetometry techniques (see Figs.~\subref{fig:fig1}{a-c}).
First, we employed x-ray scattering and found, contrary to ultrafast topological switching~\cite{buttner2021}, no evidence of enhanced short-range magnetic modulations and hence of fluctuations along the switching pathway (see insets in Figs.~\subref{fig:fig1}{a-c} and~\edf{fig:SAXS_fluctuations_viridis}).
Second, we used quantitative scanning transmission x-ray microscopy to reveal that the magnetization retains nearly its full magnitude during thermal excitation and rotates into the film plane, see~\edf{fig:max00-RB} and \hyperref[sec:SuppInfo]{Supplementary Information}.
Thus, we conclude that the dominant effect of the excitation in our material is not the emergence of a fluctuation state, but rather a spin reorientation transition into a single-domain in-plane state (Fig.~\subref{fig:fig1}{b}), followed by a spontaneous and ordered emergence of a cocoon lattice: a deterministic process that can fully be described within a micromagnetic approach.

\begin{figure*}[]
\includegraphics[width=180 mm]{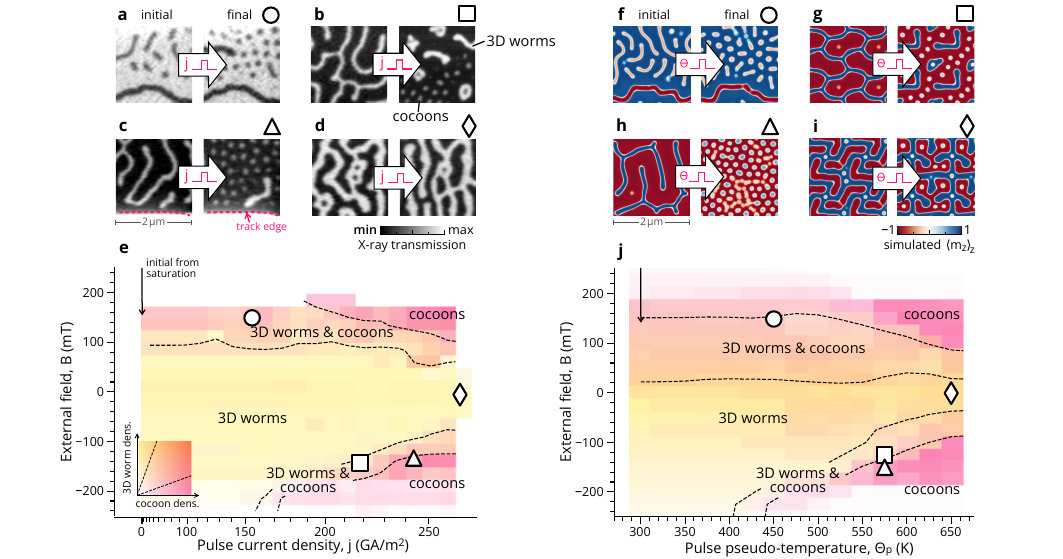}
\caption{
\textbf{Micromagnetic simulations match the experimentally observed topological transitions.}
\textbf{a}-\textbf{d,} Scanning transmission x-ray microscopy (STXM) images of representative topological transitions in a double-gradient multilayer track, here induced by \SI{40}{\ns}-long current pulses. The signal corresponds to the thickness-integrated x-ray magnetic circular dichroism.
Full black or white contrast corresponds to 3D worms, reduced contrast to cocoons.
\textbf{e,} Field-current density ($B$,$j$) phase diagram of the magnetic states produced after the current pulse (obtained from over 200 measurements). The density of cocoons and 3D worms is represented by magenta or yellow colours, respectively, as shown in the inset.
Dashed lines mark approximate phase boundaries. Symbols mark the points where the transitions in (a-d) were observed.
The $j$ axis is on a quadratic scale.
\textbf{f}-\textbf{i,} Simulated magnetic states before and after cycling of the pseudo-temperature $\Theta$ (see \hyperref[sec:Methods]{Methods}). Images display the net out-of-plane magnetization averaged along the material thickness ${\langle m_z \rangle}_z$.
\textbf{j,} Field-temperature ($B$,$\Theta_\mathrm{p}$) phase diagram of simulated transitions, analogous to the one displayed in (\textbf{e}), but for micromagnetic simulation results. 
}
\label{fig:fig2}
\end{figure*}

Indeed, zero-temperature micromagnetic simulations with time-dependent material parameters (parametrized by a pseudo-temperature parameter $\Theta$, see Fig.~\subref{fig:fig1}{d} and \hyperref[sec:Methods]{Methods}), closely reproduce the experimental results: raising $\Theta$ triggers a spin reorientation transition from the pre-existing domain state into a quasi-homogeneous state with strong in-plane components; subsequently lowering it produces a cocoon lattice (see Figs.~\subref{fig:fig1}{e-g}).
To validate our simulation framework under faster excitation conditions, we compared it with STXM observations of transitions induced by nanosecond current pulses.
As shown in Fig.~\ref{fig:fig2}, simulations and experiments exhibit excellent agreement on cocoon nucleation across all fields and current densities, both for specific exemplary transitions (Figs.~\subref{fig:fig2}{a-d,f-g}) and in the overall transition phase diagram (Figs.~\subref{fig:fig2}{e,j}).
In simulations and experiments, a dense cocoon lattice can form from a variety of initial states when subjected to sufficiently strong out-of-plane fields.
However, this agreement breaks down for 3D worms, as in experiments they stochastically appear as extended textures after strong current pulses, often emanating from sample edges (e.g. Fig.~\subref{fig:fig2}{c}).
This behaviour is characteristic of a first-order nucleation process occurring at defects, an effect that is by design absent from our fluctuation-free simulations of homogeneous materials (compare Figs.~\subref{fig:fig2}{c,h}).
The fact that simulations predict cocoon nucleation but not 3D worm formation therefore corroborates the different nucleation behaviour of the two studied textures.

\section*{Cocoon nucleation in three steps}

\begin{figure*}[]
\includegraphics[width=180 mm]{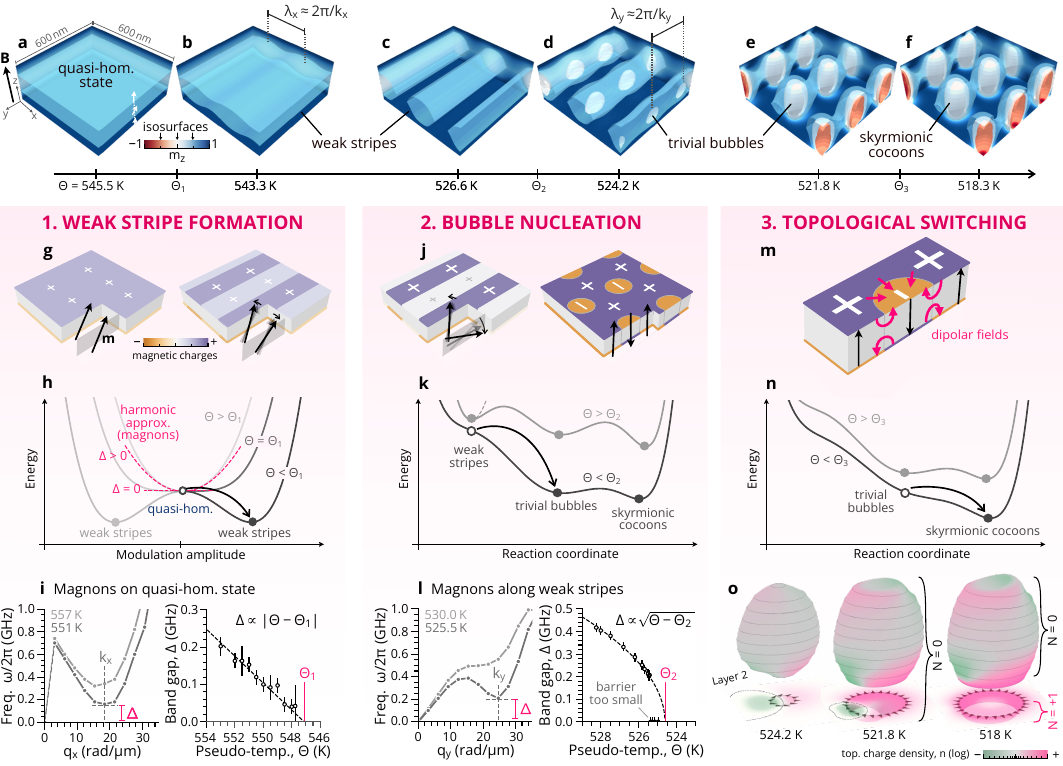}
\caption{
\textbf{The three-step process of fluctuation-free topological switching.}
\textbf{a}-\textbf{f,} Snapshots of the cocoon lattice formation process (micromagnetic simulations) upon adiabatic reduction of the pseudo-temperature $\Theta$ at $B=\SI{140}{\mT}$ in a single-gradient multilayer.
Colour represents isosurfaces of the normalized out-of-plane magnetization. Blue: $m_z=+0.5$, white: $m_z=0$, red: $m_z=-0.5$.
\textbf{g,} Sketches of surface magnetic charges during weak stripe formation.
\textbf{h,} Sketch of the energy diagram for the weak stripe transition.
\textbf{i,} Mode dispersion $\omega(\boldsymbol{q})$ of magnons propagating along the $x$ axis for two exemplary pseudo-temperatures above the first critical temperature ($\Theta >\Theta_\mathrm{1}$) and dependence of the band gap $\Delta$ as a function of $\Theta$.
The dashed line is a linear fit.
\textbf{j,} Sketches of surface magnetic charges during the weak stripe to bubble transition.
\textbf{k,} Sketch of the energy diagram for the bubble formation transition.
\textbf{l,} Mode dispersion of magnons propagating along the pre-existing weak stripe pattern displayed in (\textbf{c}) for two exemplary pseudo-temperatures above the second critical temperature ($\Theta >\Theta_\mathrm{2}$) and dependence of the band gap $\Delta$ as a function of $\Theta$. The dashed line is a fit to $\Delta \propto \sqrt{\Theta-\Theta_\mathrm{2}}$.
The points at $\Theta \approx \Theta_\mathrm{2}$ displaying an apparent $\Delta=0$ are an artifact of the method used for computing the magnon bandgap (see \hyperref[sec:Methods]{Methods}).
\textbf{m,} Sketch of the dipolar stray fields acting on a cocoon in (\textbf{e,f}).
\textbf{n,} Sketch of the energy diagram for the topological transition.
\textbf{o,} 3D representation of a single cocoon during a topological switching process. Contour: $m_z=0$ isosurface of the out-of-plane magnetization. Colour: local topological charge density $n$ on a log scale.
The 2D images below the 3D contours show the spin configuration of the domain wall that surrounds the cocoon in the second layer from the bottom, along with the local topological charge density. $N$ represents the total topological charge of each layer.
}
\label{fig:fig3}
\end{figure*}

Building on the confidence in simulations to reproduce cocoon nucleation, we use them to explore the experimentally inaccessible details of the evolution of the topological phase transition from a quasi-homogeneous state to a cocoon lattice.
Our goal is to isolate the deterministic, adiabatic mechanisms responsible for cocoon nucleation by reducing uncontrolled degrees of freedom and external sources of symmetry breaking.
We therefore restrict our analysis to the single-gradient section of the multilayer where cocoon nucleation takes place, and apply a small in-plane field on the $y$ axis to align spin textures, suppress defect-mediated transitions and enable a clear transition pathway.
In this controlled setting, we initialize the system at high pseudo-temperature $\Theta$ and under an external $B_\text{z}\SI{=140}{\mT}$ out-of-plane field, where the quasi-homogeneous configuration is the ground state.
We then simulate the transition by lowering $\Theta$ quasi-statically, mimicking adiabatic cooling until a cocoon lattice forms.
This simulation setup allows us to probe the fundamental thermodynamic response of the system and identify spontaneous transitions, without fluctuations or defects to overcome energy barriers.

The results, shown in Figs.~\subref{fig:fig3}{a-f}, reveal a three-step spontaneous nucleation process.
First, weak stripe domains emerge from the quasi-homogeneous state.
Second, the stripes collapse into a lattice of topologically trivial magnetic bubbles.
Finally, a topological transition in the outer layers converts the bubble walls from non-chiral to chiral, producing skyrmionic cocoons.
The transitions are driven by the interplay between interfacial anisotropy (favouring out-of-plane alignment), dipolar interactions (favouring magnetic charge minimization) and the external field (breaking the symmetry between ``up'' and ``down'' domains).
Their balance gives rise to successive instabilities that reshape the system without overcoming energy barriers.

\subsection*{1. Weak stripe formation}

As the system cools from the quasi-homogeneous state and anisotropy increases, the spins are driven out-of-plane and align with the external field along $+z$.
This reorientation enhances the magnetic charges (Fig.~\subref{fig:fig3}{g}), therefore raising the dipolar energy.
At a critical temperature $\Theta_\mathrm{1}$, the competition between anisotropy and dipolar interaction destabilizes the quasi-homogeneous state in favour of periodic modulations of the out-of-plane magnetization.
As a result, a weak stripe pattern emerges --- a second-order phase transition well known in low-anisotropy thin films under in-plane fields~\cite{brown1961, hubert2011, asti2007}.
In our aperiodic stacks, the weak stripes are localized in the inner lowest-anisotropy layers, while the external out-of-plane field biases the net magnetization towards the field direction.
The transition lowers the anisotropy energy while minimizing the dipolar energy cost by efficiently packing magnetic charges (see Fig.~\subref{fig:fig3}{g}, right).

The transition to weak stripes can be understood in the quasi-particle picture of magnons~\cite{grassi2022,kisielewski2023}.
In the harmonic approximation, the magnon dispersion relation $\omega(\boldsymbol{q})$ (with wave vector $\boldsymbol{q}$) is proportional to the second variation of the magnetic energy functional and thus encodes the local curvature of the energetic landscape~\cite{gurevich1996} (see Fig.~\subref{fig:fig3}{h}).
For the system to be in an equilibrium state, all magnon modes perturbing it must have positive frequencies $\omega(\boldsymbol{q})>0$, with the magnon band gap $\Delta$ corresponding to the minimum excitation energy.
When the band gap vanishes at a critical point ($\Delta \rightarrow 0$), the system becomes unstable against modulated perturbations and a transition occurs.
The scaling behaviour of the local curvature of the energy landscape (and therefore of $\Delta$) near the critical point reflects the order of the transition, independent of specific characteristics of the physical system such as the dimensionality of the sample~\cite{landau1964} (see \hyperref[sec:SuppInfo]{Supplementary Information} for details).

We compute the $\Theta$-dependent magnon dispersion on top of the quasi-homogeneous state using micromagnetic simulations (see \hyperref[sec:Methods]{Methods}).
Above a critical temperature $\Theta_\mathrm{1}$ the quasi-homogeneous state is stable, with a positive magnon bandgap $\Delta$ at a finite wave vector $\boldsymbol{q} = k_x \boldsymbol{e}_x$ (Fig.~\subref{fig:fig3}{i}).
Approaching the critical point, the band gap vanishes linearly as $\Delta \propto \abs{\Theta - \Theta_\mathrm{1}}$, indicating a second-order magnetic phase transition (see \hyperref[sec:SuppInfo]{Supplementary Information}).
The excitation of magnons at $k_x$ can then decrease the energy of the system, producing weak stripes with period $\lambda_x \approx 2\pi/\vert{k}_x\vert$, close to the inverse wave number of the softened magnon mode (see Fig.~\subref{fig:fig3}{b}).

\subsection*{2. Bubble nucleation}

Once weak stripes are established, a further reduction of $\Theta$ amplifies their out-of-plane modulation until the system reaches a tipping point at a second critical temperature $\Theta_\mathrm{2}$.
At this stage, the ``down'' domains approach in-plane orientation before being pulled into the $-z$ direction by the growing anisotropy (see~\edf{fig:Fig_CC_formation_02-Copy8}).
Rather than forming ``strong'' stripes, the pattern develops three-dimensional modulations (Fig.~\subref{fig:fig3}{d}) and relaxes into a bubble configuration (Fig.~\subref{fig:fig3}{e}).
A similar, but reversed, bubble-to-stripe transition with a three-dimensional intermediate state has been observed in FeGe crystals~\cite{yu2024}.
The stripe-to-bubble transition reduces the anisotropy energy and optimizes the dipolar energy by packing the magnetic charges efficiently into a hexagonal lattice (Fig.~\subref{fig:fig3}{j}).
The order of the transition differs fundamentally from weak stripe formation, as the modulation into a hexagonal bubble lattice cannot proceed infinitesimally once the tipping point is reached.
It is therefore a first-order transition occurring spontaneously at $\Theta_\mathrm{2}$, when the energy barrier between the two states vanishes (Fig.~\subref{fig:fig3}{k}).

In terms of magnons, the transition corresponds to the softening of a mode propagating along the stripes (i.e., with wave vector $k_y \bm{e}_y$).
As shown in Fig.~\subref{fig:fig3}{l}, the lowest-order mode along $k_y$ develops a finite minimum whose frequency $\Delta$ (directly proportional to the energy curvature) vanishes as $\Delta \propto \sqrt{\Theta - \Theta_\mathrm{2}}$.
This is consistent with a first-order phase transition where a state loses its meta-stability (see \hyperref[sec:SuppInfo]{Supplementary Information}).
Additional evidence for the first-order nature of this transition is provided by the phase coexistence between weak stripes and bubbles (see~\edf{fig:Fig_S_energies_coexistence}).

Note that the magnon spectrum exhibits a Goldstone mode at $q_y = 0$ (Fig.~\subref{fig:fig3}{l}), reflecting spontaneous symmetry breaking associated with translational degeneracy of the weak stripe phase.
During fast transitions, this symmetry breaking can generate edge dislocations in the stripe arrangement that persist as lattice defects in the final cocoon state, as seen in both experiments and simulations (Fig.~\ref{fig:fig1} and Fig.~\ref{fig:fig2}).

\subsection*{3. Topological switching}

The final stage of cocoon nucleation is a topological switching converting non-chiral bubbles into skyrmionic cocoons.
Initially, each bubble consists in every layer of two adjacent regions with opposite topological charge $\sim\pm0.5$, resembling a vortex–antivortex pair (Fig.~\subref{fig:fig3}{o}, left).
As the pseudo-temperature is decreased, the bubbles extend vertically and develop stronger out-of-plane magnetization.
During this process, their domain walls gradually reshape into chiral domain walls with Bloch line defects, especially at the surface layers of the stack (Fig.~\subref{fig:fig3}{o}, centre).
At a third critical pseudo-temperature $\Theta_\mathrm{3}$, a topological transition occurs.
A chiral domain alignment emerges in the bottom layer, forming a N\'eel cap with integer topological charge and conferring a skyrmionic character to the spin texture.
Other layers undergo similar transitions at different temperatures and fields, and an opposite N\'eel cap can form also at the top of the stack~\cite{grelier2022}.

The fact that topological switching starts at the outer layers aligns with the known cocoon structure, featuring a non-chiral core and homochiral domain walls at the surfaces~\cite{grelier2022}.
This heterogeneous configuration is stabilized by internal stray fields, which act as effective chiral forces that promote opposite chirality near the two surfaces of the multilayer (see Fig.~\subref{fig:fig3}{m}).
In our material, the DMI further promotes the counter-clockwise chirality, favouring the N\'eel cap at the bottom of the stack~\cite{grelier2022, savchenko2023}.
Chiral stray fields are ubiquitous to all magnetic materials, even in systems with inversion symmetry, with strength that can surpass that of DMI fields~\cite{desautels2019, weber2025}.
Therefore, as we show in the next section, the here-proposed topological switching mechanism is not exclusive to our graded multilayers, but can instead occur in a wide range of magnetic thin films.

\section*{Universality of the three-step pathway}

\begin{figure*}[]
\includegraphics[width=180 mm]{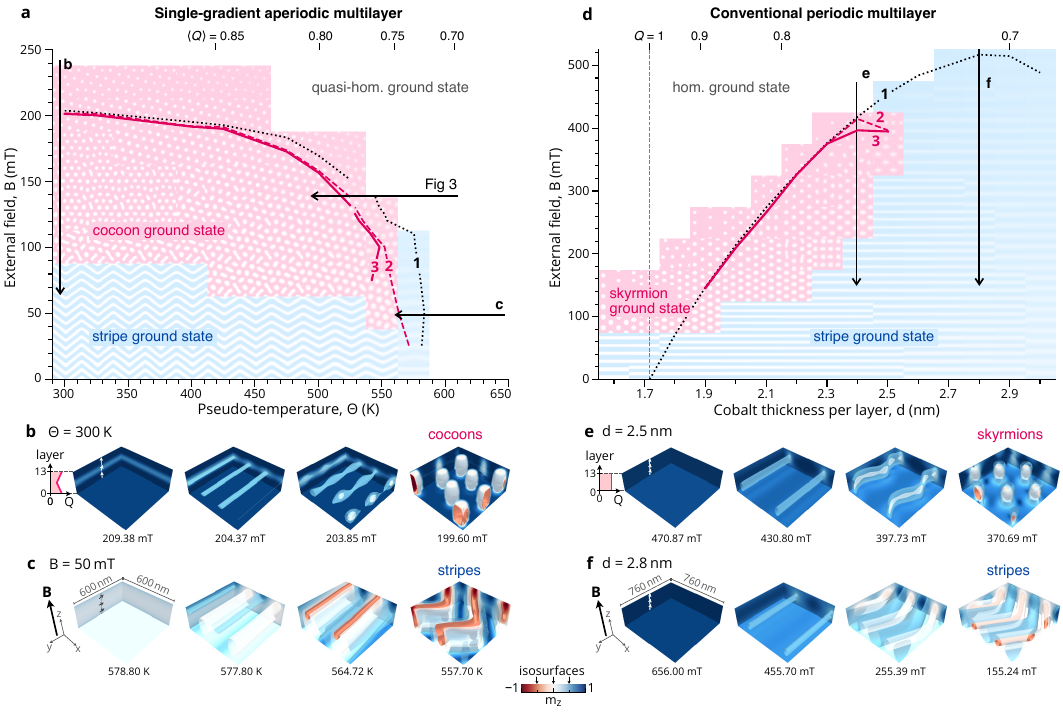}
\caption{
\textbf{Universality of fluctuation-free topological switching.}
\textbf{a,} Phase space diagram (zero-temperature simulations) of ground states in a single-gradient multilayer as a function of out-of-plane field $B$ and pseudo-temperature $\Theta$.
White: quasi-homogeneous state, magenta: cocoon lattice, blue: zig-zag stripes~\cite{aitoukaci2023}.
The $z$-averaged quality factor $\langle Q \rangle$ of the material is indicated on the top axis.
Arrows indicate representative paths taken through the phase space.
The dotted, dashed and solid lines mark the weak stripe formation, bubble nucleation and topological switching transitions, respectively, when coming from the quasi-homogeneous phase.
\textbf{b,c,} Snapshots of states along the paths marked in (\textbf{a}), displayed as isosurfaces of the normalized out-of-plane magnetization $m_z$. Blue: $m_z=+0.5$, white: $m_z=0$, red: $m_z=-0.5$.
\textbf{d,} Phase space diagram (zero-temperature simulations) of ground states in a conventional [Pt$\mid$Co($d$)$\mid$Al]\textsubscript{\texttimes 13} periodic magnetic multilayer.
White: homogeneous state, magenta: skyrmion lattice, blue: stripes.
Lines: see (\textbf{a}).
\textbf{e,f,} Snapshots of states along the paths marked in (\textbf{d}).
} 
\label{fig:fig4}
\end{figure*}

The three-step nucleation pathway provides indeed a general mechanism for the spontaneous formation of cocoons, as shown by micromagnetic simulations in Figs.~\subref{fig:fig4}{a–c}.
Across broad field and temperature ranges, and under both field and temperature cycling, cocoons consistently emerge through the same sequence of transitions, each playing a distinct symmetry breaking role at an associated critical line in the $B$–$\Theta$ phase space.
First, a second-order phase transition destabilizes the quasi-homogeneous state, breaking translational symmetry along the $x$ axis and generating weak stripes.
Second, a modulational instability deforms these stripes along their length, breaking translational symmetry along the $y$ axis.
The outcome of this second transition depends on the ground state of the system at the corresponding field and temperature $(B,\Theta)$.
At finite fields, the degeneracy between ``up’’ and ``down’’ domains is lifted, and the transition produces a lattice of magnetic bubbles (Fig.~\subref{fig:fig4}{b}); in the absence of a field, the ground state is instead a stripe phase (Fig.~\subref{fig:fig4}{c}), consistent with literature on weak stripes~\cite{brown1961, hubert2011, asti2007, aitoukaci2023} and on ferroic patterns after thermal quenching~\cite{horstmann2025}.
Finally, a topological switching event transforms the bubble lattice into skyrmionic textures.

To test the generality of this pathway beyond graded multilayers, we simulated periodic (non-graded) stacks with fixed cobalt thickness and well-defined quality factor $Q$.
The results are shown in Figs.~\subref{fig:fig4}{d–f}, where spontaneous stripe and skyrmion nucleation can be observed during field-cycling from the saturated state via the same sequence of modulational instabilities.
These simpler systems clarify the conditions required for spontaneous topological lattice nucleation.
A quality factor $Q < 1$ is necessary for weak stripes to destabilize the homogeneous state (occurring at the black dotted line in Fig.~\subref{fig:fig4}{d}) and to allow the formed stripes to eventually fragment into bubbles.
This condition can be understood within the domain wall energy model~\cite{cape1971}: in high-anisotropy materials, the positive domain wall energy creates an energy barrier preventing both the nucleation of magnetic domains and stripe-to-bubble transitions~\cite{hubert2011, cape1971, moon2024}, while for $Q<1$ this energy barrier is not necessarily present.
Additionally, the bubble lattice must be energetically favoured during the transition, excluding the very low $Q \ll 1$ regime where stripes are the ground state (Fig.~\subref{fig:fig4}{f}).
In a restricted range of $Q$ both criteria are met and spontaneous skyrmion nucleation is observed (Fig.~\subref{fig:fig4}{e}).

\section*{Discussion}

The softening of linear quasi-particle excitations as a precursor of phase transitions is a common feature of nonlinear systems and can also be observed, for example, close to crystalline phase transitions~\cite{zhao2024}, in quantum gases~\cite{mottl2012}, or dipolar superfluids~\cite{biagioni2022}.
Here, we have shown that such pathways can also drive topological transitions in magnetic systems.

\begin{figure*}[]
\includegraphics[width=89 mm]{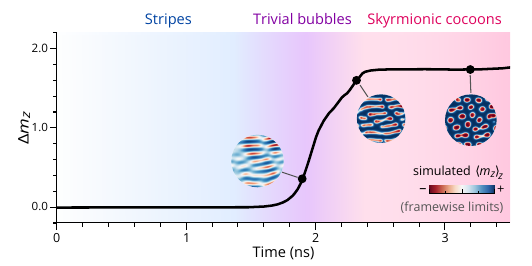}
\caption{
\textbf{Speed of fluctuation-free topological switching (simulation).}
The plot shows the modulation amplitude ($\Delta m_z = {\langle m_z \rangle}_z^\text{max} - {\langle m_z \rangle}_z^\text{min}$) of the $z$-averaged out-of-plane magnetization, i.e., the metric that corresponds to the total intensity in a time-resolved SAXS experiment. The initial state (before time zero) is a quasi-homogeneous ground state at $\Theta=\SI{650}{\K}$. Time is measured starting from the instant lowering of the pseudo-temperature $\Theta$ is to \SI{300}{\K}.
The shaded areas are a guide to the eye, marking the succession of magnetic states.
Insets display exemplary states during the nucleation process.
}
\label{fig:fig5}
\end{figure*}

The proposed fluctuation-free mechanism contrasts with conventional models of skyrmion formation that rely on thermal fluctuations~\cite{buttner2021}.
However, the two processes are not mutually exclusive and may both contribute to cocoon formation.
In the static measurements discussed here, the persistence of a finite order parameter and the agreement with simulations indicate that the instability-driven, fluctuation-free pathway is dominant.
Nonetheless, in other experimental conditions and for ultrafast excitation stimuli, this may change.
Figure~\ref{fig:fig5} shows the simulated speed of cocoon nucleation after instantaneous cooling from a quasi-homogeneous state.
While the transition from weak stripe domains to skyrmionic cocoons occurs within a few hundred picoseconds, a timescale comparable to laser-induced skyrmion nucleation, the formation of the weak-stripe precursor can extend over one nanosecond.
At ultrafast timescales, thermal fluctuations may therefore play a more decisive role, and the balance between the two nucleation pathways remains an open question.
Future time-resolved experiments may disentangle these two contributions: SAXS could distinguish between fluctuations and weak stripes immediately after excitation, while time-resolved magnetometry could establish whether a spin reorientation transition occurs even during ultrafast excitation processes. 

\section*{Conclusion}

Our results establish a general pathway for the nucleation of topological magnetic domains, proceeding through a sequence of spontaneous modulational instabilities.
Our work also provides design principles for engineering fluctuation-free formation of topological states in magnetic thin films. The key requirement is an effective quality factor $Q < 1$, which allows weak stripes to destabilize the homogeneous state, combined with an out-of-plane field favouring the stripe-to-bubble transition.
In high-anisotropy multilayers, a transient anisotropy reduction by nonthermal laser switching~\cite{stupakiewicz2017}, ionic gating~\cite{dacamarasantaclaragomes2024}, or strain~\cite{kong2025} could produce the right conditions for this pathway, enabling the deterministic nucleation of stable topological textures in spintronic devices.
More broadly, the principle of topological texture nucleation via modulational instabilities may extend beyond magnetic cocoons and skyrmions to other textures, and even to non-magnetic ordered media such as ferroelectric films hosting polar skyrmions~\cite{wang2023}.

\vspace{15mm}

\textbf{Acknowledgements:}
We thank the Helmholtz-Zentrum Berlin für Materialien und Energie GmbH for the allocation of synchrotron radiation beamtime.
Part of the experiments were performed on the SEXTANTS beamline at SOLEIL Synchrotron, France (proposal number 20241803).
We acknowledge the MAX IV Laboratory for beamtime on the SoftiMAX beamline under proposal 20241001.
We acknowledge DESY (Hamburg, Germany), a member of the Helmholtz Association HGF, for the provision of experimental facilities.
Parts of this research were carried out at P04.
Beamtime was allocated for proposals I-20220515, I-20221288, I-20230608.
We acknowledge the Paul Scherrer Institut, Villigen, Switzerland for provision of synchrotron radiation beamtime at the beamline PolLux of the SLS.
We acknowledge the use of the Physical properties laboratory, which is part of the CoreLab ``Quantum Materials'' operated by HZB.
We thank Yara E. Mahboub for her participation to beamtime experiments.

\textbf{Funding:} We acknowledge funding from the Helmholtz Young Investigator Group Program through Project No. VH-NG-1520, from the Deutsche Forschungsgemeinschaft (DFG, German Research Foundation) through project number 505818345 (Topo3D) and 49254781 (TRR 360) subproject C02, from the Leibniz-Gemeinschaft through project number K162/2018 (OptiSPIN) and from the French National Research Agency (ANR) through project numbers ANR-22-EXSP-0002 (PEPR SPIN CHIREX) and ANR-22-CE92-0082 (Topo3D).
L.K. acknowledges funding by the Radboud Excellence Initiative and by the Initiative and Networking Fund of the Helmholtz Association.
T.A.B. acknowledges funding from SNI and the European Regional Development Fund (ERDF).
Research conducted at MAX IV, a Swedish national user facility, is supported by Vetenskapsrådet (Swedish Research Council, VR) under contract 2018-07152, Vinnova (Swedish Governmental Agency for Innovation Systems) under contract 2018-04969 and Formas under contract 2019-02496.
Development of SOPHIE was supported by the Swiss Nanoscience Institute (SNI).
The PolLux end station was financed by the German Ministerium für Bildung und Forschung (BMBF) through ErUM-Pro contracts 05K16WED, 05K19WE2 and 05K22WE2.

\textbf{Author Contributions}:
R.Ba. and F.B. conceptualized the work.
R.Ba., M.G., S.Wit., C.M.G., N.R. and S.P. prepared the samples. 
R.Ba., K.L., K.P.J., M.G., S.Wit.,D.M., L-M.K., C.K., T.K., R.Bo., M.P., S.Win., M.W., M.S., J.F., N.J., S.F., J.R., T.A.B., H.P., J.R. and N.R. acquired the data at x-ray experiments.
K.P. acquired the SQUID data. 
R.Ba. analyzed the experimental data.
L.K. provided the theoretical framework.
R.Ba. performed micromagnetic modeling with \textsc{MuMax3}, supported by L.K., M.G and K.L.
R.Ba., L.K., K.L., N.J., V.C., B.P., J.H.M., N.R. and F.B. discussed and interpreted the results.
R.Ba. and L.K. visualized the results and wrote the original draft.
S.E., V.C., B.P., J.H.M, N.R. and F.B. supervised the project.
All authors reviewed and commented on the manuscript.

\textbf{Competing Interests}:
The authors declare no competing interests.

\textbf{Data availability}:
The raw data, source data and analysis code for all figures are available at
https://zenodo.org/doi/10.5281/zenodo.18004538 (ref.~\cite{battistelli2025b}).

\section*{Methods}\label{sec:Methods}
\subsection*{Sample preparation}

The composition of our Ta(\SI{5}{\nm})$\mid$[Pt(\SI{3}{\nm})$\mid$Co($d$)$\mid$Al(\SI{1.4}{\nm})]\textsubscript{\texttimes N}$\mid$Pt(\SI{3}{\nm}) AMLs is analogous to the one detailed in previous works~\cite{grelier2022, grelier2023a, grelier2023}.

Single-gradient stacks are composed of $N_\text{SG}$ repeats in which the cobalt thickness $d$ varies with the layer $i$ as:
\begin{equation}
\begin{cases}
d_i = d_0 + S\cdot i & \text{for } i\leq N_\text{SG}/2\\
d_i = d_0 + S\cdot (N_\text{SG}-1-i) & \text{for } i> N_\text{SG}/2
\end{cases}
\end{equation}
where $d_0$ is the cobalt thickness of the first ($i=0$) and last ($i=N_\text{SG}-1$) layers, and $S$ is the thickness step between two consecutive layers.
$S=\SI{0.1}{\nm}$ for every sample studied or simulated in this work.

The double-gradient stacks studied for this project (Figs.~\subref{fig:fig1}{a-c},\ref{fig:fig2}) are composed of a conventional high-anisotropy multilayer stack of $N_\text{ML}\SI{=15}{}$ repeats with cobalt thickness $d_\text{ML}\SI{=0.9}{\nm}$ in the middle, flanked by two single-gradient sections at the top and bottom ($N_\text{SG}\SI{=13}{}$, $d_0 = \SI{1.6}{\nm}$), for a total of 41 repeats.
To enable current excitation, we structured the magnetic film into racetracks of well-defined length and width with dimensions of \SI{5}{\micro\meter}\texttimes\SI{20}{\micro\meter} via UV lithography.
The SAXS measurements displayed in the insets in Figs.~\subref{fig:fig1}{a-c} were obtained on a single-gradient multilayer ($N_\text{SG}\SI{=13}{}$, $d_0 = \SI{1.6}{\nm}$), fabricated into a \SI{150}{\micro\meter}-wide magnetic track to enable continuous current excitation.
The STXM magnetometry measurements mentioned in the main text and displayed in~\edf{fig:max00-RB} were obtained on a single-gradient multilayer ($N_\text{SG}\SI{=13}{}$, $d_0 = \SI{2.0}{\nm}$), fabricated into a \SI{5}{\micro\meter}\texttimes\SI{20}{\micro\meter} racetrack.
All samples imaged in x-ray experiments were deposited via argon-ion-assisted DC magnetron sputtering on \SI{200}{\nm} thick silicon-nitride membranes to allow for x-ray transmission.
For superconducting quantum interference device (SQUID) measurements we used a periodic Ta(\SI{5}{\nm})$\mid$ [Pt(\SI{3}{\nm})$\mid$Co(\SI{1.5}{\nm})$\mid$Al(\SI{1.4}{\nm})]\textsubscript{\texttimes 3}$\mid$Pt(\SI{3}{\nm}) multilayer grown on a silicon wafer.

\subsection*{X-ray transmission experiments}

The microscopy images displayed in Figs.~\ref{fig:fig1}, \ref{fig:fig2}, \subref{fig:Fig_transitions_means2}{a-c} and~\subedf{fig:FigDC_heating}{b,c} were acquired via scanning transmission x-ray microscopy (STXM) at the MAXYMUS endstation of the BESSY II electron storage ring, operated by the Helmholtz-Zentrum Berlin für Materialien und Energie GmbH.
Small-angle x-ray scattering (SAXS) experiments (insets in Figs.~\subref{fig:fig1}{a-c} and~\subedf{fig:SAXS_fluctuations_viridis}{e-i}) were performed at the SEXTANTS endstation of the SOLEIL synchotron.
The scattering patterns displayed in this paper were acquired at \SI{85}{\mT}.
Patterns obtained from a saturated sample were subtracted to remove spurious background components such as stray light.

X-ray ptychography of samples heated via direct currents (\subedf{fig:FigDC_heating}{d} and~\subedf{fig:FIG_SG_heating_vs_T_fromMAXIV}{a}) was performed at the SoftiMAX beamline of the MAX IV Laboratory (Lund, Sweden) with the Soft X-ray Ptychography Highly Integrated Endstation (SOPHIE)~\cite{butcher2025}.
The distance between the sample and the detector was \SI{96}{\mm}.
A single-photon counting iLGAD EIGER detector with 512x512 pixels (\SI{75}{\micro\m} pixel size)~\cite{baruffaldi2025} was used to record diffraction patterns at \SI{100}{\nm} step interval from a \SI{900}{\nm} FWHM x-ray beam shaped by a line doubled Ir Fresnel zone plate (diameter: \SI{500}{\micro\m}, outer zone width: \SI{20}{\nm}).
The ptychographic datasets were reconstructed with the PtychoShelves software package~\cite{wakonig2020}, using the difference-map~\cite{thibault2008} and the maximum-likelihood refinement~\cite{odstrcil2018} methods with three probe modes~\cite{thibault2013}.

FTH microscopy experiments on infrared laser-induced cocoon nucleation were performed at the P04 beamline of the PETRA III electron storage ring (\subedf{fig:Fig_transitions_means2}{d}), while preliminary FTH measurements (not shown in this work) were performed at the PolLux beamline of the SLS synchrotron facility.

To obtain magnetic contrast, all experiments were performed using circularly polarized light at the L\textsubscript{3}-edge of cobalt (wavelength \SI{1.59}{\nm}) in order to exploit the x-ray circular magnetic dichroism effect (XMCD).
STXM images and SAXS datasets were acquired recording only images with one circular polarization, while FTH images, ptychography reconstructions and STXM magnetometry data were acquired recording datasets with both polarizations.

A variable out-of-plane magnetic field $B$ was applied by a set of four rotating permanent magnets at MAXYMUS and SEXTANTS, by an electromagnet at P04 and by a permanent magnet at SoftiMAX.

\subsection*{Experimental phase diagram computation}

For the experimental phase diagram shown in Fig.~\subref{fig:fig2}{e}, around 200 different transitions were observed with STXM, for a total of around 400 images (one for the initial and one for the final state). See \hyperref[sec:SuppInfo]{Supplementary Information} for more of the observed final states.
Transitions were observed at eight different external magnetic fields ($B=\SI{200}{}$, \SI{140}{}, \SI{100}{}, \SI{0}{}, \SI{-50}{}, \SI{-100}{}, \SI{-140}{}, \SI{-200}{\mT}) and for electrical current pulses with a \SI{40}{\ns} pulse width and current density amplitudes $j$ ranging from \SI{0}{} to \SI{263}{\giga\A/\m^2}.
The amplitude and duration of the current pulses were measured using an oscilloscope.
In order to compute the phase diagram, we analyzed every final state obtained after the transition was complete, and computed two relevant metrics.
First, we counted the number of cocoons present in the final state and second, we measured the area occupied by 3D worms.
This gives us a qualitative phase diagram dense on the $j$ axis, but very discrete on the $B$ axis, with just the eight field points mentioned above.
However, including just the externally applied field disregards the transient Oersted fields produced by the current pulses.
Indeed, we are able to reliably observe Oersted field effects in the transitions, promoting the annihilation of 3D worms in different regions of the magnetic track depending on the direction of the applied current.
These effects were not noticeable with continuous current, as the current densities used (and therefore the resulting Oersted fields) are roughly 3 orders of magnitude smaller.

To take into account Oersted field effects, the phase space diagram reported in the main text includes the net out-of-plane component of the Oersted fields acting in the material calculated from the known dimensions of the magnetic track and the current density flowing through it~\cite{zeissler2020}, averaged along the vertical axis for all layers of the stack.
The in-plane components of the Oersted field and their non-uniform distribution along the $z$ axis were neglected.
For each detected cocoon, we computed the Oersted field acting on it based on its position in the magnetic track and the current density of the pulse, and added this field to the static external field.
The same process was followed to establish the total effective fields acting on a particular portion of a 3D worm.
Therefore, the phase diagram in the main text describes the final state of the material as a function of the current density and total effective magnetic field during the current pulse.

\subsection*{Micromagnetic parameters evaluation}

The exchange stiffness $A=\SI{18}{\pico\J/\m}$ and the interfacial DMI constant $D_\mathrm{s}=\SI{-2.34}{\pico\J/\m}$ at room temperature were taken from previous work on similar AMLs~\cite{grelier2022, grelier2023}.
The temperature-dependent values of the saturation magnetization $M_\mathrm{s}(T)$ and interfacial anisotropy $K_\mathrm{s}(T)$ (Fig.~\subref{fig:fig1}{d}) were obtained by measuring SQUID in-plane and out-of-plane hysteresis loops of a Ta(\SI{5}{\nm}) $\mid$  [Pt(\SI{3}{\nm})$\mid$Co(\SI{1.5}{\nm})$\mid$Al(\SI{1.4}{\nm})]\textsubscript{\texttimes 3} $\mid$ Pt(\SI{3}{\nm}) multilayer at temperatures ranging from \SI{300}{\K} to \SI{650}{\K}, measuring every \SI{50}{\K}.
The in-plane and out-of-plane loops were both normalized by the area of the measured samples, and spurious experimental backgrounds were subtracted.
The temperature-dependent saturation magnetization $M_\mathrm{s}(T)$ was then directly obtained from the magnetization of the sample at high fields, while the effective anisotropy $K_\mathrm{eff}(T)$ was obtained 
from the out-of-plane and in-plane hysteresis curves using the method employed in~\cite{buttner2013}.
The interfacial uniaxial anisotropy $K_\mathrm{s}(T)$ was then computed as $K_\mathrm{s}(T)= d \cdot \left( K_\mathrm{eff}(T)+\frac{\mu_0}{2} M_\mathrm{s}^2(T)\right)$, where $d$ is the thickness of the magnetic layer and $\mu_0$ is the magnetic permeability constant.
See~\edf{fig:SQUID_vs_T} for more details.
The obtained temperature-dependent quantities were then fitted using the equations:
\begin{equation}
 M_\mathrm{s}(T) = M_\mathrm{s}^0 \cdot \left [ 1- \left( \frac{T}{T_\mathrm{c}} \right)^{e_m} \right]
 \label{eq:MsT}
\end{equation}
where $M_\mathrm{s}^0$ is the saturation magnetization at zero temperature, $T_\mathrm{c}$ is the Curie temperature of the material and $e_m$ an phenomenological parameter~\cite{schlotter2018, lee2017}, and
\begin{equation}
 K_\mathrm{s}(T) = K_\mathrm{s}^0 \cdot \left( 1-b \frac{T}{T_\mathrm{c}} \right)\cdot\left(\frac{M_\mathrm{s}(T)}{M_\mathrm{s}^0}\right)^{e_k}
 \label{eq:KuT}
\end{equation} 
where $K_\mathrm{s}^0$ is the uniaxial anisotropy at zero temperature, $b$ is a phenomenological parameter related to the thermal expansion coefficient of the material and $e_k$ another phenomenological parameter~\cite{lee2017}.
The fit parameters are $M_\mathrm{s}^0$\SI{=1.248}{\mega\A\m^{-1}}, $K_\mathrm{s}^0$\SI{=1.728}{\milli\J\m^{-2}}, $T_\mathrm{c}$\SI{=919.2}{\K}, $e_m$\SI{=3.135}{}, $b$\SI{=3.135e-9}{}, $e_k$\SI{=2.67}{}, with the resulting curves shown in Fig.~\subref{fig:fig1}{d}.
For the simulations in Fig.~\ref{fig:fig3} and Fig.~\ref{fig:fig4}, slightly different fitting parameters were used resulting from a different background subtracting procedure ($M_\mathrm{s}^0$\SI{=1.284}{\mega\A\m^{-1}}, $K_\mathrm{s}^0$\SI{=1.820}{\milli\J\m^{-2}}, $T_\mathrm{c}$\SI{=915.9}{\K}, $e_m$\SI{=3.140}{}, $b$\SI{=8.496e-17}{}, $e_k$\SI{=2.73}{}, see blue line in~\edf{fig:SQUID_vs_T}).
These slightly different parameters do not significantly affect the simulations, and all the phenomenological behaviours are identical. None of the conclusions reached in this work is affected by the choice of any specific set of micromagnetic parameters.

The temperature-dependent interfacial DMI constant was obtained by assuming a scaling with the saturation magnetization as:
\begin{equation}
 D_\mathrm{s}(T) \propto M_\mathrm{s}^5(T)
\end{equation}
Similar scaling relations are present in literature for other PtCo-type multilayers, with exponents ranging from $\sim$1.8 to $\sim$5~\cite{schlotter2018,zhou2020,alshammari2021,ham2022}.

In simulations, it appears that the most important parameter relevant to the spin reorientation transition is the quality factor $Q$.
Changing or removing the temperature dependence of $D$ does not alter the observed phenomenon significantly: at sufficient temperatures a quasi-homogeneous state in the single-gradient multilayer is always observed, and restoring the original interfacial anisotropy provokes cocoon nucleation.

In this present work, the temperature dependence of the exchange stiffness $A$ was ignored.
Nonetheless, we performed micromagnetic simulations where the exchange stiffness was assumed to follow the same temperature dependence as the interfacial anisotropy, as found by Asti and colleagues~\cite{asti2007}, which lead to results similar to our approach.

\subsection*{Micromagnetic simulations}

The micromagnetic simulations were performed using the solver \textsc{MuMax3} (version 3.10)~\cite{vansteenkiste2014}, using a procedure similar to~\cite{grelier2022}.
We modeled a slab with \text{N}\textsubscript{x}\texttimes\text{N}\textsubscript{y}\texttimes\text{N}\textsubscript{z} cells of size \SI{4}{\nm}\texttimes\SI{4}{\nm}\texttimes\SI{2.133}{\nm}.
To take into account the presence of a non-magnetic buffer between the Co layers (\SI{3}{\nm} of Pt and \SI{1.4}{\nm} of Al), we divided each trilayer into three layers, one magnetic and two non-magnetic ones.
The number of vertical cells N\textsubscript{z} varied depending the simulated stack (N\textsubscript{z}\SI{=121}{} for the double-gradient in Figs.~\ref{fig:fig1},\ref{fig:fig2}, N\textsubscript{z}\SI{=37}{} for the single-gradient and multilayer stacks in Figs.~\ref{fig:fig3},\ref{fig:fig4},\ref{fig:fig5}).
The number of lateral cells N\textsubscript{x,y} varied depending on computational constraint and time considerations (N\textsubscript{x,y}\SI{=512}{} for Fig.~\ref{fig:fig1}, Figs.~\subref{fig:fig2}{f,h} and Fig.~\ref{fig:fig3}, N\textsubscript{x,y}\SI{=256}{} for Figs.~\subref{fig:fig2}{g,i,j} and Fig.~\ref{fig:fig4}).
The structure of the simulated samples are $N_\text{SG}=13$, $d_0=\SI{1.6}{\nm}$, $N_\text{ML}=15$, $d_\text{ML}=\SI{1.0}{\nm}$ for the double-gradient multilayers in  Fig.~\ref{fig:fig1},\ref{fig:fig2} and $N_\text{SG}=13$, $d_0=\SI{2.0}{\nm}$ for the single-gradient multilayers in Fig.~\ref{fig:fig3},\subref{fig:fig4}{a-c},\ref{fig:fig5}.

The magnetic parameters were rescaled using the effective medium model~\cite{woo2016} as in the following.
A scaling factor $s=d/p$ was defined, where $d$ and $p\SI{=2.133}{\nm}$ are the thicknesses of the magnetic and simulation layers, respectively.
Due to the aperiodic nature of the cobalt thickness $d$ in each individual layer of the stack, $s$ is layer dependent, conferring a gradient on the third axis to the micromagnetic properties of the studied material.
The saturation magnetization and exchange stiffness were multiplied by $s$, while the uniaxial anisotropy was rescaled following the formula $K_\mathrm{u}=K_\mathrm{s}/d \cdot s - \mu_0 M_\mathrm{s}^2(s-s^2)$ and the DMI as $D = D_\mathrm{s} / d \cdot s$.
The simulations of spin-reorientation induced transitions shown in Figs.~\subref{fig:fig1}{e-g} and~\subref{fig:fig2}{f-j} were performed by initializing the simulation from a state obtained via field cycling and changing $K_\mathrm{s}$, $M_\mathrm{s}$ and $D_\mathrm{s}$ as a function of the pseudo-temperature parameter $\Theta$.
The system was relaxed, obtaining the high-temperature transient states (e.g. Fig.~\subref{fig:fig1}{e}). Then, the micromagnetic parameters were restored to their original values and the system relaxed into its final state (e.g. Fig.~\subref{fig:fig1}{g}, right side).

For the micromagnetic simulation study of a single-gradient stack shown in Fig.~\ref{fig:fig3}, we first obtained the succession of stable states starting from a high pseudo-temperature $\Theta=\SI{590}{\K}$, for which the quasi-homogeneous state is the stable state, and then slowly reduced $\Theta$ at a rate of \SI{0.0225}{\K/\ns} while the system evolved.

The states obtained with the slow temperature sweeping were also used as a starting point for the calculations of the band dispersion plots shown in Figs.~\subref{fig:fig3}{o,p}.
Each state was first relaxed, and then perturbed for \SI{100}{\ns} with a field $\boldsymbol{B}(\boldsymbol{r},t)=(dB_x(\boldsymbol{r},t), 0, B_z)$, where the modulation added to the external field is:
\begin{equation}
 dB_x(\boldsymbol{r},t) = G \frac{\sin(k_c(r-5\pi/k_c))}{k_c(r-5\pi/k_c)}\frac{\sin(\omega_c (t-10\pi/\omega_c))}{\omega_c (t-10\pi/\omega_c)}
\end{equation}
where $r$ is the distance of each point from the $x$ or $y$ axis (in the case of the weak stripe (Fig.~\subref{fig:fig3}{o}) and cocoon (Fig.~\subref{fig:fig3}{p}) nucleation, respectively), $t$ is the time, $G=\SI{1}{\mT}$, $k_c=\frac{\pi}{4}\SI{}{\nm^{-1}}$ and $\omega_c=4\pi \SI{}{\giga\Hz}$.
The magnetic state perturbed by $dB_x$ was saved every \SI{0.1}{\ns}, and the $m_y$ component of the magnetization on the central layer along a single line of pixels on the $x$ and $y$ axes (for Fig.~\subref{fig:fig3}{o} and Fig.~\subref{fig:fig3}{p}, respectively) was chosen, producing the 2D time- and space-resolved plots of the magnetization $m_y(x/y,t)$ (\subedf{fig:magnondispersion}{a-d}).
The band dispersion plot $m_y(q_{x/y}, \omega/2\pi)$ were then calculated via a 2D Fourier transform (\subedf{fig:magnondispersion}{e-h}), from which the dispersion of the lowest energy mode and the magnon bangap $\Delta$ were extracted.
The points displaying an apparent null bandgap ($\Delta=0$) close to the second transition critical point $\Theta_\mathrm{2}$ are an artifact of the method used. For $\Theta \approx \Theta_\mathrm{2}$ the energy barrier separating the stripe and bubble state is further reduced, and any small perturbation is enough to overcome it and provoke a cocoon nucleation event.
The fit of the dependence of $\Delta$ as a function of $\Theta$ are $\Delta = c_1 \left| \Theta - \Theta_\mathrm{1} \right| $ ($c_1\SI{=0.033}{\giga\Hz/K}$, $\Theta_1\SI{=546.73}{\K}$) for the first transition and $\Delta = c_2 \sqrt{ \Theta - \Theta_\mathrm{2} } $ ($c_2\SI{=0.220}{\giga\Hz/\sqrt{\K}}$, $\Theta_2\SI{=524.57}{\K}$) for the second transition.
The critical temperature $\Theta_3$\SI{=521.64}{\K} for the third transition was obtained by monitoring changes in the topological charge of the texture.

The ground state phase diagrams shown in Figs.~\subref{fig:fig4}{a,d} were computed for each couple of temperature/field $(\Theta,B)$ and cobalt thickness/field $(d,B)$ values by initializing the magnetic state from multiple distinct magnetic configuration (i.e. starting with a certain number of aligned stripes or skyrmions, or a homogenous state), letting the system relax and recording its total energy. The lowest energy state was chosen as the ground state of the system under those conditions.
The three critical lines were extracted by initializing the system as a homogeneous state and then slowly varying the conditions (i.e. decreasing $B$ or $\Theta$ for Fig.~\subref{fig:fig4}{a} and decreasing $B$ for Fig.~\subref{fig:fig4}{d}), monitoring whether and when the various transitions occurred.
In this case, no magnon band analysis was performed and the critical fields and temperatures were evaluated manually, observing at what point the state developed stripes, bubble-like modulations or a non-trivial topology, respectively for the first, second and third critical lines.
The magnetic states produced during these slow field/temperature sweeps were also used for the exemplary states shown in Figs.~\subref{fig:fig4}{b,c,e,f}.

The topological charge density $n(x,y)=\frac{1}{4\pi}\boldsymbol{m}\cdot\left(\frac{\partial\boldsymbol{m}}{\partial x}\times\frac{\partial\boldsymbol{m}}{\partial y}\right)$ was computed using the Berg and L\"uscher method~\cite{berg1981}.
To obtain the net topological charge of a magnetic texture, $n(x,y)$ was integrated over the area occupied by the texture of interest as $N= \iint n(x,y)\,dx\,dy$, which equates to a sum $N= \sum_{i,j} n_{i,j}$ for a discretized simulated texture.

\beginesupplementary
\label{sec:SuppInfo}

\section*{Observation of the spin reorientation transition with STXM magnetometry}

To obtain the STXM magnetometry data shown in \edf{fig:max00-RB} we measured the field-dependent XMCD transmission intensity (which is proportional to the magnetization of the sample along the light propagation direction) of a single-gradient multilayer sample illuminated using a defocused x-ray beam (approximately \SI{4}{\micro\m}$\times$\SI{4}{\micro\m} in size) to obtain an average information over a large area:
\begin{equation}
\text{XMCD}=\frac{I^+ - I^-}{I^+ + I^-}
\label{eq:stxm0}
\end{equation}
where $I^+$ and $I^-$ are the intensities recorded for left- and right-circularly polarized light, respectively.

The sample was tilted with respect to the light propagation axis $\hat{k}$ by \SI{30}{\degree} to gain sensitivity to the magnetization in-plane component $m_\mathrm{x}$.
The external magnetic field $B$ was applied at angles of $\alpha =\SI{0}{\degree}, \SI{60}{\degree}, \SI{90}{\degree}$ with respect to the light propagation axis.
The measured XMCD contrast was normalized to the saturated signal at room temperature, yielding a measurement equal to the projection of the net magnetization onto the light propagation axis $\hat{k}$:
\begin{equation}
\text{XMCD} \approx m_\text{k}(T) = \frac{M_\text{k}(T)}{M_\text{s}(\SI{300}{K})}
\label{eq:stxm1}
\end{equation}
where $M_k$ is a linear combination of the in-plane and out-of-plane components of the magnetization:
\begin{equation}
M_\text{k} = \cos(30^\circ) M_\text{x} + \sin(30^\circ) M_\text{z},
\label{eq:stxm2}
\end{equation}
For sufficiently strong external fields, the magnetization can be considered uniform and lying on the $xz$-plane ($\mathbf{M} = (M_x, 0, M_z)$), so that its projection on the light propagation axis can be expressed as:
\begin{equation}
M_\text{k}(T)= M_\text{s}(T) \cos{\theta}
\label{eq:stxm3}
\end{equation}
where $\theta$ is the angle between the magnetization and $\hat{k}$.

From Eq.~\ref{eq:stxm1} and~\ref{eq:stxm3}, we obtain a direct relationship between the recorded signal, the saturation magnetization at the temperature of the experiment and $\theta$:
\begin{equation}
\text{XMCD} \approx \frac{M_\text{s}(T)}{M_\text{s}(\SI{300}{K})} \cos{\theta}
\label{eq:stxm4}
\end{equation}

\edf{fig:max00-RB} shows hysteresis loops of $\text{XMCD}$ vs. the field $B$ obtained at room temperature and under heating by a current density $j=\SI{10.8}{\giga\A/\m^2}$, sufficient to fully suppress magnetic domains (\subedf{fig:max00-RB}{e}).
The observed hysteresis loops are fully compatible with a spin reorientation transition resulting in a uniform magnetization with strong in-plane components.

Under small magnetic fields, the heated sample displays a magnetic signal $\text{XMCD} \approx 0.5$, compatible with fully in-plane alignment of the magnetization and a saturation magnetizion equal to the one at room temperature ($\theta=\SI{60}{\degree}$, $M_\text{s}(T)= M_\text{s}(\SI{300}{K})$).
For fields along the sample plane ($\alpha\SI{=60}{\degree}$), the signal remains stable at $\SI{0.5}{}$ as the field increases, confirming the in-plane alignment of the magnetization.
For tilted fields, after the in-plane alignment at small fields ($\theta \approx \SI{60}{\degree}$), further increasing the field produces an increase or decrease of the magnetic signal, respectively in the case of fields applied parallel ($\alpha\SI{=0}{\degree}$) and perpendicular ($\alpha\SI{=90}{\degree}$) to the light propagation axis.
This corresponds to the magnetization progressively aligning to the external field ($\theta \rightarrow \alpha$), towards or away from the light propagation axis for $\alpha\SI{=0}{\degree}$ and $\alpha\SI{=90}{\degree}$, respectively.

Since the normalized XMCD signal under resistive heating reaches a value very close to $\SI{0.5}{}$, we can conclude that the ratio $\frac{M_\text{s}(T)}{M_\text{s}(\SI{300}{K})} \approx 1$.
Therefore, $M_s(T)$ remains essentially unchanged during heating.

\section*{Derivation of the frequency scaling of the magnon bandgap for the modulational instabilities}

In this Section, we want to establish a connection between the frequency scaling of the magnon bandgap $\Delta$ and the nature of a modulational magnetic phase transition.

\subsection*{Connection between magnon frequency and phase stability}

An essential prerequisite for understanding the scaling behavior of the magnon bandgap near a magnetic phase transition is its relation to the micromagnetic energy $\mathcal{G}(\boldsymbol{m},\mu)$ of the magnetic system driven through a phase transition controlled by some control parameter $\mu$.
In the present context, this parameter may represent either the applied magnetic field $B_z$ or a quasi-temperature $\Theta$.
The energy functional $\mathcal{G}$ contains various material parameters, such as the uniaxial anisotropy $K(\mu)$, whose change with the control parameter can lead to bifurcations in the dynamics of a magnetic system.
Such bifurcations can involve instabilities of formerly stationary magnetic configurations $\boldsymbol{m}_0$ and can show characteristics of thermodynamic phase transitions.

To study the critical behavior of a micromagnetic system close to such an instability, we expand its energy around a stable configuration $\boldsymbol{m}_0$ with respect to small variations $\delta\boldsymbol{m}\perp\boldsymbol{m}_0$ as
\begin{equation}\label{eq:variation-expansion}
    \mathcal{G}(\boldsymbol{m}_0+\delta\boldsymbol{m},\mu) = \mathcal{G}(\boldsymbol{m}_0,\delta\boldsymbol{m},\mu) + \frac{1}{2!}\delta^2 \mathcal{G}(\boldsymbol{m}_0,\delta\boldsymbol{m},\mu) + \frac{1}{3!}\delta^3 \mathcal{G}(\boldsymbol{m}_0,\delta\boldsymbol{m},\mu) + \frac{1}{4!}\delta^4 \mathcal{G}(\boldsymbol{m}_0,\delta\boldsymbol{m},\mu) + ...
\end{equation}
where $\delta^{(n)}\mathcal{G}_\mu$ denotes the $n$th variation. Here, we have already set the first variation $\delta\mathcal{G}=0$ once we are interested in the critical behavior where $\boldsymbol{m}_0$ is still stationary. The second variation can be written as 
\begin{equation}\label{eq:second-variation}
    \delta^2 \mathcal{G}(\boldsymbol{m}_0,\delta\boldsymbol{m}_0;\mu) = \int\limits_V\mathrm{d}V\, \delta\boldsymbol{m}\cdot\vu{H}_0(\mu)\cdot\delta\boldsymbol{m}
\end{equation}
with $\vu{H}_0$ being the $\mu$-dependent Hessian of the micromagnetic energy at $\boldsymbol{m}_0$ with respect to admissible variations. From Eq.~\eqref{eq:second-variation} it is clear that $\boldsymbol{m}_0$ is stable to all admissible $\delta\boldsymbol{m}$ if $\vu{H}_0$ has only positive eigenvalues. In this case, these eigenvalues $\omega_\nu(\mu)$ (with a mode index $\nu$) correspond to the linear frequencies of the magnetic normal modes (magnon excitations) around $\boldsymbol{m}_0$ and are, just like $\mathcal{G}$, dependent on the control parameter $\mu$. Within nonlinear spin-wave theory \cite{tyberkevych2020,perna2022}, the micromagnetic energy is expressed in terms of the Fourier amplitudes $\psi_\nu(t)$ of these normal modes by $\delta\boldsymbol{m}=\sum_\nu \psi_\nu(t)\boldsymbol{m}_\nu(\boldsymbol{r}) + \mathrm{c.c.}$ with $\boldsymbol{m}_\nu$ being the complex mode profiles obtained from the linearized Landau-Lifshitz equation.  

Instabilities arise when the lowest magnon frequency approaches zero, that is, when the magnon bandgap $\Delta = \min(\omega_\nu)$ and, therefore, the second variation $\delta^2\mathcal{G}$ into the direction of this mode, vanishes. Depending on the nature of the instability, $\Delta$ can become negative or even imaginary, and the corresponding soft mode can \textit{grow} into a noncollinear spin texture \cite{kisielewskiWavesPatternsSpin2023}. Hence, to study the critical behavior near such a transition, it is sufficient to consider only the lowest energy magnon mode, $\psi_\nu \equiv \psi$, whose frequency defines the bandgap. The expansion Eq.~\eqref{eq:variation-expansion} can then be simplified by projecting it only onto the lowest energy magnon mode. After suitable normalization of the corresponding mode profile, one obtains 
\begin{equation}
    \mathcal{G} = \mathcal{G}_0 + \Delta(\mu)\abs{\psi}^2 + V(\mu) \psi^*\abs{\psi}^2  + W(\mu) \abs{\psi}^4 + \mathrm{c.c.} + ...
\end{equation}
with $\mathcal{G}_0=\mathcal{G}(\bm{m}_0)$, and $V(\mu)$ and $W(\mu)$ being three- and four-magnon coefficients that can also be $\mu$-dependent. 

\subsection*{Weak-stripe formation (Second-order transition)}

\begin{figure}[h!]
    \centering
    \includegraphics{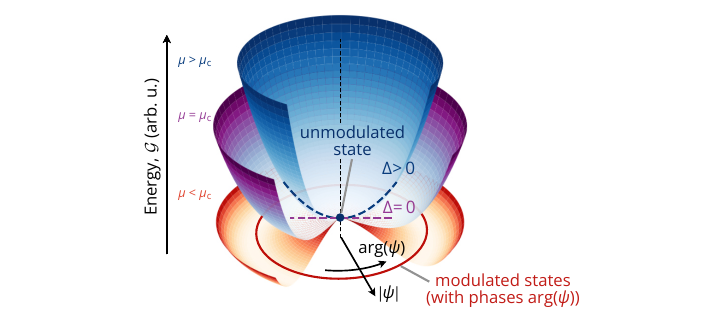}
    \caption{Schematic representation of the effective potential during a second-order modulational instability accompanied by the softening of a magnon mode with amplitude $\psi$. During the transition, the magnon bandgap $\Delta$ on top of the unmodulated state $\abs{\psi}=0$ becomes negative, giving rise to a degenerate set of stable modulated states at some non-zero $\abs{\psi}$.}
    \label{fig:second-order}
\end{figure}

The nucleation of weak stripe patterns can be interpreted as a continuous (second-order) phase transition from an unmodulated to a modulated state whereby the magnitude of the dominant Fourier component of this modulation with wave vector $\boldsymbol{k}$ departs from zero as soon as $\mu < \mu_\mathrm{c}$. In terms of the normal-mode picture described in the previous section, such an instability occurs when a magnon mode with a non-zero wave-vector on top of the homogeneous state becomes soft. The effective energy expansion in terms of the softening mode close to the critical point $\mu_\mathrm{c}$ reads
\begin{equation}
    \mathcal{G}(\psi,\mu) = \mathcal{G}_0 + \Delta(\mu)\abs{\psi}^2 + W\abs{\psi}^4 + ...
\end{equation}
with a four-wave coefficient $W$ that we take as constant in the control parameter $\mu$ for simplicity. All odd coefficients vanish by symmetry \cite{grassiHiggsGoldstoneSpinwave2022}. Here, "..." includes higher-order terms in $\psi$ and possible but irrelevant contributions of higher-frequency magnon modes. Before the phase transition $\mu > \mu_\mathrm{c}$, the gap $\Delta(\mu)$ is positive and the equilibrium is at the unmodulated state $\psi_0=0$ [see Fig.~\ref{fig:second-order}]. As $\mu$ crosses $\mu_\mathrm{c}$, the harmonic coefficient $\Delta$ changes its sign and the point $\psi_0$ loses its stability through a pitchfork bifurcation, leading to a smooth increase of the modulation magnitude $\abs{\psi}$ [see Fig.~\ref{fig:second-order}]. To change its sign at $\mu_\mathrm{c}$, the bandgap $\Delta$ has to be odd function in $\mu-\mu_\mathrm{c}$. Therefore, close to the critical point, we find in leading order
\begin{equation}
    \Delta \propto (\mu-\mu_\mathrm{c}).
\end{equation}
Thus, close to a second-order phase transition, the magnon bandgap $\Delta$ is linear in $\mu-\mu_\mathrm{c}$. 

\subsection*{Bubble nucleation (first-order transition)}

\begin{figure}[h!]
    \centering
    \includegraphics{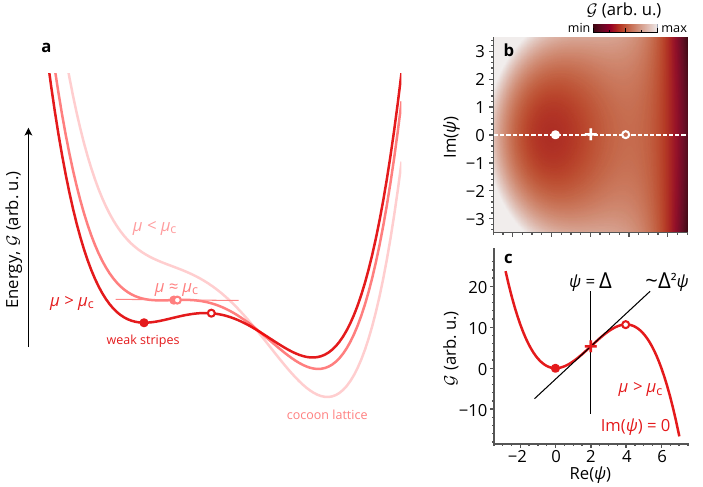}
    \caption{(a) Schematic representation of a double-well potential during a first-order phase transition. The full dot shows the stable stripe state, the hollow dot the unstable transition state. (b) Micromagnetic energy in leading order according to Eq.~\eqref{eq:free-energy-first-order}. (c) Energy along the real axis of the mode amplitude $\psi$ with the slope of the energy annotated at the inflection point.}
    \label{fig:scaling_first_order}
\end{figure}

Let us now determine the frequency scaling of the magnon bandgap, $\Delta$, along the stripe domains near the point of bubble nucleation. Similar to stripe formation, this transition proceeds via a modulational instability; however, in contrast to the continuous (second-order) stripe instability, it is \textit{discontinuous} (first order), as evidenced by the presence of hysteresis. Consequently, the energy landscape can be schematically represented as an antisymmetric double-well potential. At the critical (spinodal) point, the stripe state merges with the transition state located at the top of the energy barrier (see Fig.~\ref{fig:scaling_first_order}a) through a saddle-node bifurcation. This coalescence is accompanied by the vanishing (eventual sign change) of the slope at the inflection point between the stripe and the transition state (see Fig.~\ref{fig:scaling_first_order}a,c).  

To investigate how the magnon bandgap $\Delta$ scales during this transition, we expand the micromagnetic energy around the stable stripe state in terms of the amplitude of the softening magnon mode $\psi$, including the lowest-order three-magnon contributions as
\begin{equation}\label{eq:free-energy-first-order}
    \mathcal{G}(\psi,\mu) = \Delta(\mu)\,|\psi|^2 - \frac{1}{6}V\,\psi^*|\psi|^2 + \mathrm{c.c.} + \ldots
\end{equation}
Here, ``c.c.'' denotes the complex conjugate of the preceding term, and $V$ is a (possibly complex) three-magnon coupling coefficient, which we set to $V=1$ for simplicity. With this phase convention, both the stripe and the transition states lie on the real axis [Fig.~\ref{fig:scaling_first_order}b]. One can readily show that the micromagnetic energy exhibits an inflection point along the real axis at $\mathrm{Re}(\psi) = \Delta$, where the first derivative along this axis is equal to $\Delta^2$ [Fig.~\ref{fig:scaling_first_order}c]. At the spinodal point $\mu = \mu_\mathrm{c}$, this derivative must change sign and therefore behaves as an odd function of $(\mu - \mu_\mathrm{c})$. Hence, to leading order,
\begin{equation}
    \Delta^2 \propto \mu - \mu_\mathrm{c},
\end{equation}
which implies that the magnon bandgap closes following the scaling law
\begin{equation}
    \Delta \propto \sqrt{\mu - \mu_\mathrm{c}}.
\end{equation}
Thus, near a first-order transition, the lowest magnon mode softens with the square-root behavior characteristic of a saddle-node bifurcation.

\section*{Experimental phasespace derivation}

\begin{figure*} [hbt!]
\includegraphics [width=180 mm]{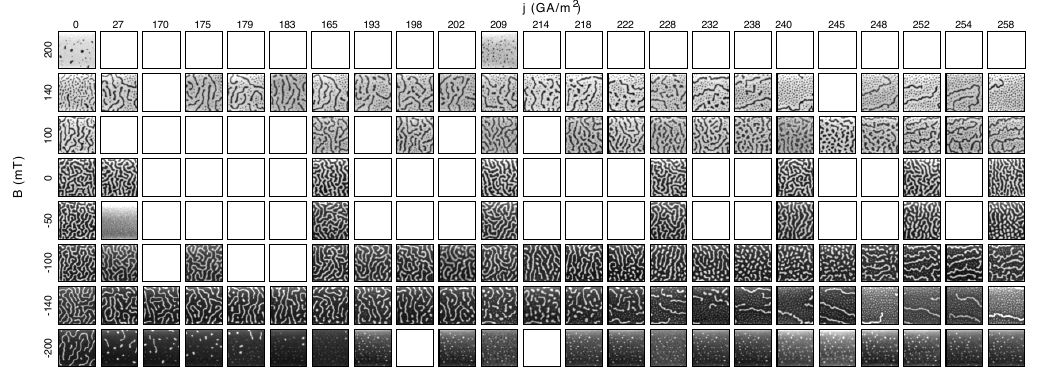}
\caption{
\textbf{Experimental nanosecond current pulse-induced transitions in a double-gradient magnetic multilayer.}
Examples of images acquired after excitation with a \SI{40}{\ns} current pulse, for different external fields $B$ and applied current densities $j$.
The asymmetry in the orientation of the isolated 3D worms at $\SI{\pm140}{\mT}$ can be explained via spin-orbit torque effects of the current pulse affecting the 3D worm propagation in the medium.
These images were employed to compute the phase diagram shown in Fig.~\subref{fig:fig2}{e}.
Sample structure is (\textbf{a-c}) $N_\text{SG}=13$, $d_0=\SI{1.6}{\nm}$, $S=\SI{0.1}{\nm}$, $N_\text{ML}=15$, $d_\text{ML}=\SI{0.9}{\nm}$.
}
\label{fig:FigDGpulses_SI_exp_phasespace}
\end{figure*}

For the experimental phase diagram shown in Fig.~\subref{fig:fig2}{e}, around 200 transitions were observed at eight different external magnetic fields ($B=\SI{200}{}$, \SI{140}{}, \SI{100}{}, \SI{0}{}, \SI{-50}{}, \SI{-100}{}, \SI{-140}{}, \SI{-200}{\mT}), for electrical current pulses with a \SI{40}{\ns} pulse width and current density amplitudes $j$ ranging from \SI{0}{} to \SI{263}{\giga\A/\m^2}.
Some exemplary final states of these transitions are shown in Fig.~\ref{fig:FigDGpulses_SI_exp_phasespace}.

The out-of-plane component of the Oersted field produced by the current pulse in every point of the magnetic track was computed as~\cite{zeissler2020}:

\begin{equation}
H_\text{z}=\frac{I}{8 \pi w t} \int_{-w/2}^{+w/2}\int_{-t/2}^{+t/2}  \left( \frac{z'-z}{ {(x'-x)}^2 + {(z'-z)}2}  \right)  \,dx \,dz,
\end{equation}

where $I$ is the current flowing through the magnetic track, $w=\SI{5}{\micro\meter}$ is the track width, $t$ is the multilayer thickness, $x$ and $z$ are the coordinates of each point with respect to the center of the track (in this coordinate system, the track lies along the $y$ axis).

The in-plane components of the Oersted field were neglected, and only the net out-of-plane component averaged over the sample thickness was considered.
This net out-of-plane Oersted field, different for every observed skyrmionic cocoon and for every point of the observed 3D worms, was added to the externally applied out-of-plane field in order to compute the total effective field acting on a particular texture during the transition.
For each combination of current density of the electrical pulse and total effective field, the probability of nucleating skyrmionic cocoons and 3D worms were computed, and used to compute the colors in the phase phase diagram displayed in ~\subref{fig:fig2}{e}.
The results of missing combinations of current pulses and fields (white spaces in Fig.~\ref{fig:FigDGpulses_SI_exp_phasespace} were extrapolated from neighbouring points in the phase diagram.

\begin{figure*} [hbt!]
\includegraphics [width=180 mm]{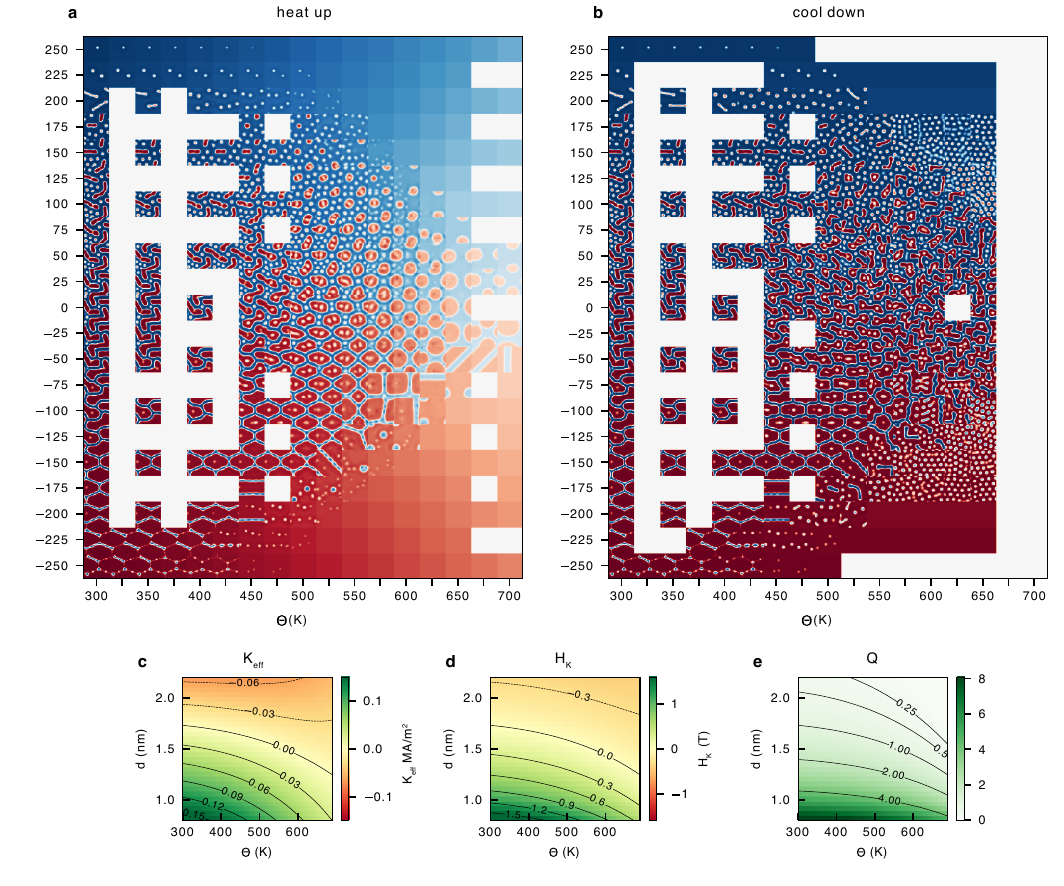}
\caption{
\textbf{Simulations of temperature-induced transitions in a double-gradient magnetic multilayer.}
States obtained with thermal excitations, increasing the pseudo-temperature $\Theta$ to a maximum $\Theta_p$ for \SI{40}{\ns} and then restoring it to \SI{300}{\K}, for different external fields $B$ and maximum temperatures $\Theta_p$.
\textbf{a} States at high temperatures $\Theta = \Theta_p$.
\textbf{b} Final states after cooling back to $\Theta =\SI{300}{\K}$.
These states were used to compute the phase diagram shown in Fig.~\subref{fig:fig2}{j}.
\textbf{c,d,e} Effective anisotropy $K_\text{eff}=K_\text{u}-\frac{\mu_0 M_\text{s}^2}{2}$, perpendicular magnetic anisotropy field $H_\text{K}=\frac{2K_\text{eff}}{M_\text{s}}$ and quality factor $Q=\frac{2K_\text{u}}{\mu_0 M_\text{s}^2}$ of a layer as a function of cobalt layer thickness $d$ and quasi-temperature $\Theta$.
}
\label{fig:mumax_vs_T}
\end{figure*}

\section*{Mumax phasespace derivation}

To obtain the phase diagram displayed in Fig.~\subref{fig:fig2}{j} of the main text we computed micromagnetic simulations of a double-gradient multilayer sample.
First, we simulated an out-of-plane hysteresis loop of the sample at room temperature (Fig.~\subref{fig:mumax_vs_T}{a}, $\Theta=\SI{300}{\K}$), obtaining a magnetic state of the sample every $\SI{25}{\K}$.
For each field, we obtained ``heated'' states by increasing $\Theta$ and letting the system evolve for $\SI{20}{\ns}$ and relax.
States at various temperatures up to $\SI{650}{\K}$ were simulated (Fig.~\subref{fig:mumax_vs_T}{a}).
Then, the ``cooled'' states were simulated from the ``heated'' ones by lowering $\Theta$ back to $\SI{300}{\K}$, letting the system evolve for $\SI{20}{\ns}$ and relax (Fig.~\subref{fig:mumax_vs_T}{b}).
For each combination of field and maximum pseudo-temperature reached, the probability of nucleating skyrmionic cocoons (i.e., the cocoon density) and the area occupied by the 3D worms were computed, and used to compute the colors in the phase phase diagram displayed in Fig.~\subref{fig:fig2}{j}.
For some combination of fields and temperatures, the simulations were not performed (white spaces in Figs.~\subref{fig:mumax_vs_T}{a,b} and the cocoon density and 3D worms area were extrapolated from neighbouring points in the phase diagram.

\clearpage

\bibliography{bib}

\clearpage

\beginextendeddata

\begin{figure*}[h]
\includegraphics[width=18 cm]{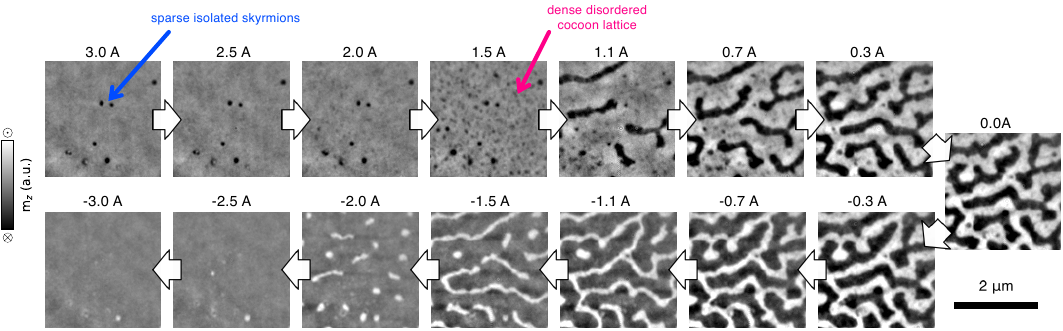}
\caption{
\textbf{Spontaneous cocoon nucleation during field cycling}
Out-of-plane hysteresis loop of a double-gradient multilayer, acquired with FTH.
Skyrmions (darker spots) nucleate in a few sparse centers, likely associated with material defects. Cocoons (lighter spots) nucleate in dense lattices.
The labels report the current applied to the electromagnet, proportional to the out-of-plane field applied to the sample.
An exact calibration of the externally applied field is not available for this measurement.
The sample structure is $d_0\SI{=1.6}{\nm}$, $N_\text{SG}=13$ layers, $S\SI{=0.1}{\nm}$, $N_\text{ML}=15$, $d_\text{ML}\SI{=1.5}{\nm}$.
}
\label{fig:EXTFIG1}
\end{figure*}
\newpage

\begin{figure*}[h]
\includegraphics[width=18 cm]{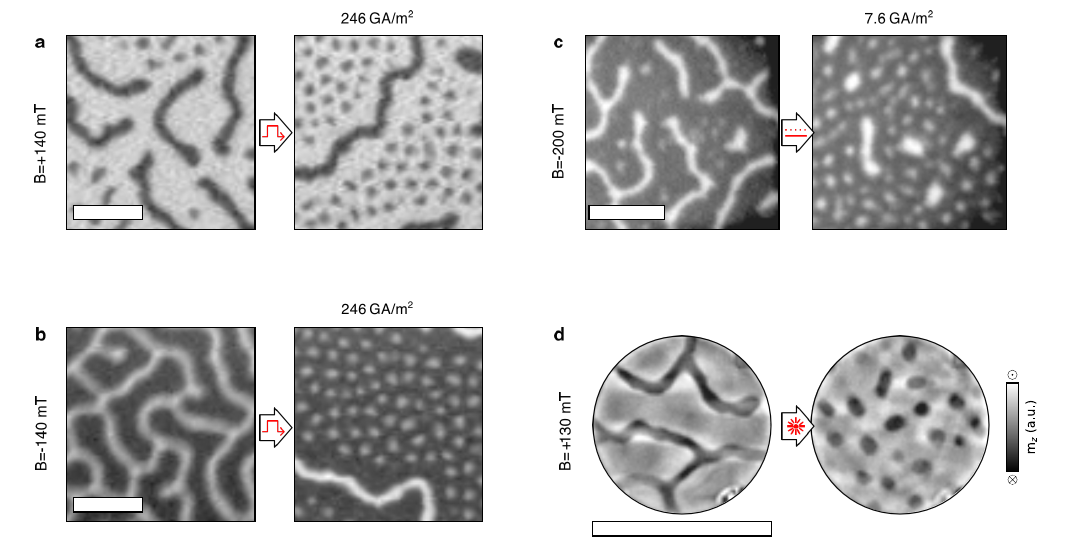}
\caption{
\textbf{Topological switching induced by a variety of excitations.}
\textbf{a-d,} X-ray microscopy images of magnetic states before and after excitation, produced via (\textbf{a,b}) \SI{40}{\ns} current pulses, (\textbf{c}) direct current heating and (\textbf{d}) femtosecond infrared laser pulses.
All images show the thickness-averaged x-ray magnetic circular dichroism contrast in double-gradient magnetic multilayers. Sample structures are (\textbf{a-c}) $N_\text{SG}=13$, $d_0=\SI{1.6}{\nm}$, $N_\text{ML}=15$, $d_\text{ML}=\SI{1.0}{\nm}$ and (\textbf{d}) $N_\text{SG}=13$, $d_0=\SI{1.6}{\nm}$, $N_\text{ML}=15$, $d_\text{ML}=\SI{0.9}{\nm}$.
(\textbf{a-c}) are STXM images, (\textbf{d}) are FTH reconstructions.
The inhomogeneous magnetic contrast in (d) is due to artifacts of the FTH microscopy technique.
White scalebars are \SI{1}{\micro\meter}.
}
\label{fig:Fig_transitions_means2}
\end{figure*}
\newpage

\begin{figure*}[h]
\includegraphics[width=18 cm]{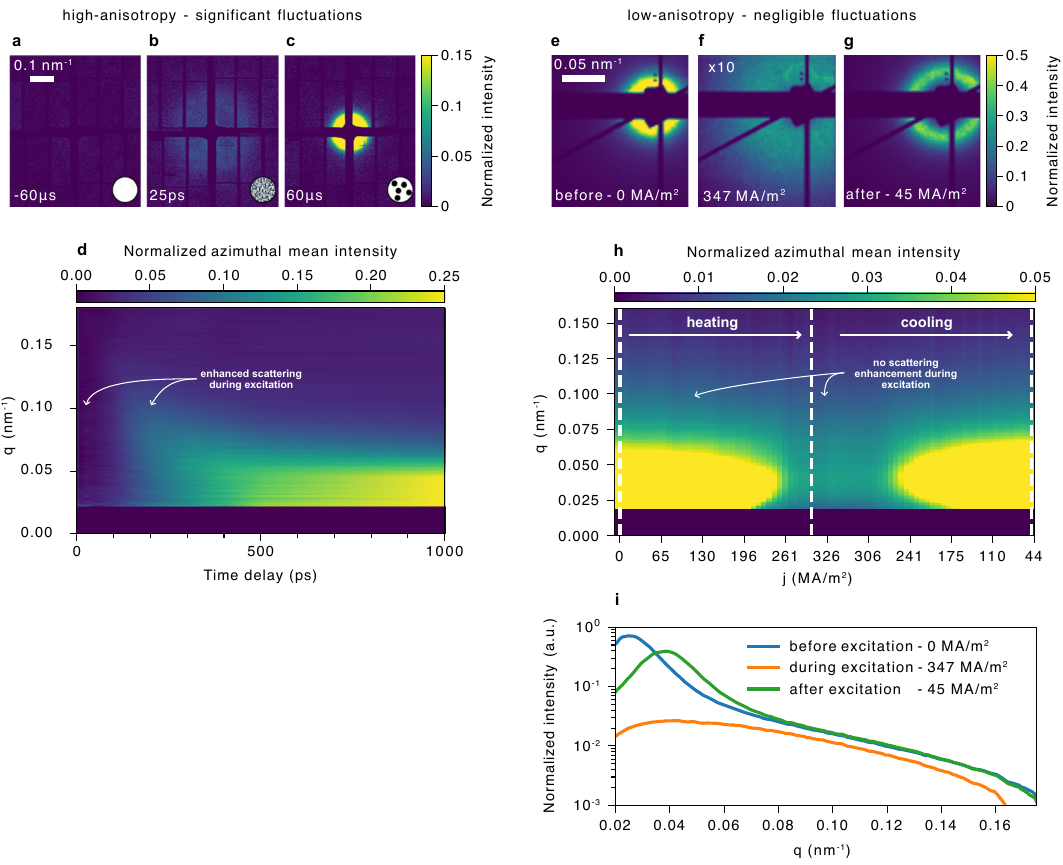}
\caption{
\textbf{Comparison between fluctuation-mediated skyrmion nucleation and fluctuation-free cocoon nucleation.}
\textbf{a-d} Experimental SAXS study of femtosecond infrared laser-induced skyrmion nucleation in a high-anisotropy multilayer, reproduced from Ref.~\cite{buttner2021}.
\textbf{a-c}, Example SAXS patterns before (\textbf{a}), during (\textbf{b}) and after (\textbf{c}) skyrmion nucleation. The delay between the respective x-ray pulse and the infrared pulse is indicated.
Insets: sketches of the corresponding real-space spin textures.
\textbf{d}, q-dependent scattering (azimuthal average) of the transient state as a function of pump–probe delay. The intensity I at larger q values corresponds to higher spatial gradients in real space. The dashed line indicates the fitted peak positions.
During excitation, the scattering at high spatial frequencies is enhanced, indicating an increased presence of short range magnetic fluctuations.
\textbf{e-f}, SAXS patterns measured after field cycling at \SI{85}{\milli\tesla} (\textbf{e}), during continuous current excitation (\textbf{f}, current $I=\SI{4.25}{\milli\A}$, current density $j=\SI{347}{\mega\A\m^{-2}}$), and after the current density has been gradually decreased (\textbf{g}, current $I=\SI{0.55}{\milli\A}$, current density $j=\SI{45}{\mega\A\m^{-2}}$). The signal in (\textbf{f}) has been multiplied by a factor 10 for clarity.
The current applied to the sample was slowly ramped up and down for the heating/cooling cycle.
\textbf{h}, Azimuthal averages of the SAXS intensity as a function of applied current density $j$ during slow heating and cooling.
\textbf{i}, Azimuthal averages of the SAXS intensity for the patterns displayed in (\textbf{e-g}).
Unlike in (\textbf{a-d}), the scattering at high-q during excitation is not enhanced, proving that the provided continuous current excitation does not induce magnetic fluctuations.
The diffuse scattering observed in (\textbf{f}) may instead be due to static modulations of the magnetization such as those observed in~\subedf{fig:max00-RB}{e,f} and~\subedf{fig:FIG_SG_heating_vs_T_fromMAXIV}{a}.
Sample parameters: $N_\text{SG}=13$, $d_0=\SI{1.6}{\nm}$.
All intensities were normalized to the largest value of the final states.
}
\label{fig:SAXS_fluctuations_viridis}
\end{figure*}
\newpage

\begin{figure*}[h]
\includegraphics[width=170 mm]{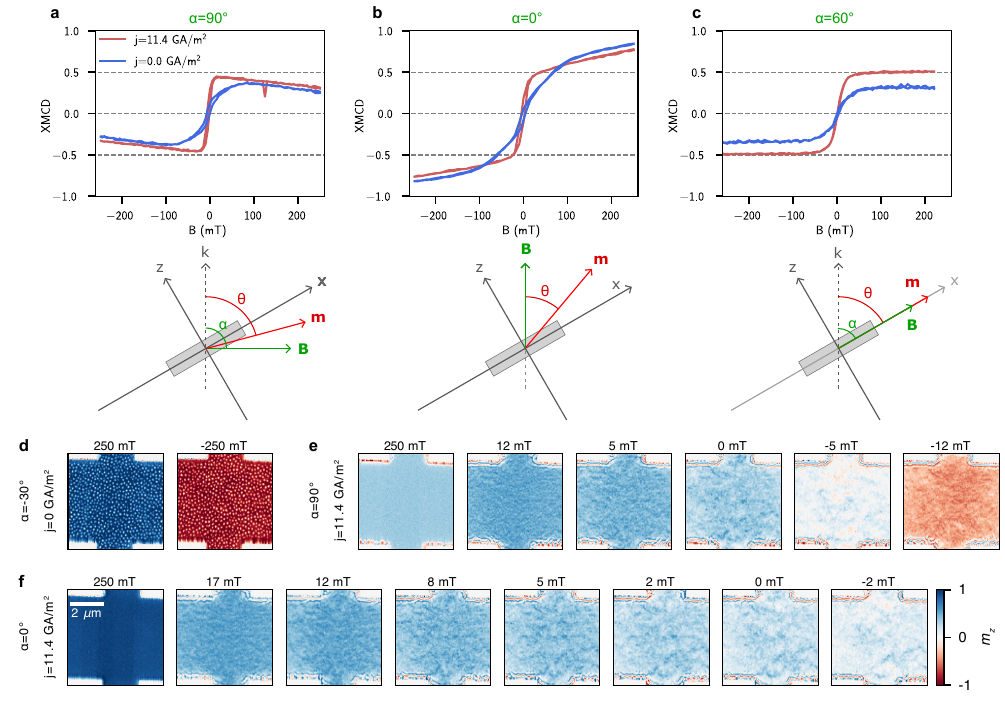}
\caption{
\textbf{Spin reorientation transition in a single-gradient multilayer induced by direct current resistive heating.}
\textbf{a-c,} Hysteresis loops of the magnetic signal $\text{XMCD}$, proportional to the component of the magnetization parallel to the xrays $m_\text{k}$, both at room temperature and while heated with a direct current density $j=\SI{11.4}{\giga\A/\m^2}$.
Measurements were obtained by measuring the x-ray transmission of the sample in a tilted geometry, under an external magnetic field $B$ tilted by (\textbf{a}) $\alpha=\SI{90}{\degree}$, (\textbf{b}) $\alpha=\SI{0}{\degree}$ and (\textbf{c}) $\alpha=\SI{60}{\degree}$ with respect to the light propagation axis $k$.
\textbf{d} STXM images acquired at the maximum available out-of-plane fields ($B\SI{=250}{\mT}$, $\alpha=\SI{-30}{\degree}$) without applying a current to the sample. These images were used to calculate the maximum XMCD contrast used to normalize all other images and hysteresis loops.
\textbf{e} STXM images acquired during hysteresis loops with $\alpha=\SI{90}{\degree}$ and an applied current density $j=\SI{11.4}{\giga\A/\m^2}$.
\textbf{f} STXM images acquired during hysteresis loops with $\alpha=\SI{0}{\degree}$ and an applied current density $j=\SI{11.4}{\giga\A/\m^2}$.
Sample structure is $N_\text{SG}=13$, $d_0=\SI{2.0}{\nm}$.
See \hyperref[sec:SuppInfo]{Supplementary Information} for details.
}
\label{fig:max00-RB}
\end{figure*}
\newpage

\begin{figure*}[h]
\includegraphics[width=90 mm]{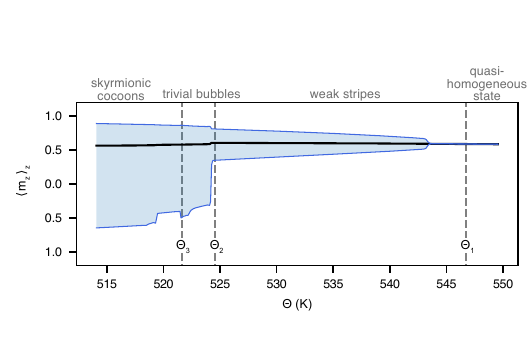}
\caption{
\textbf{Evolution of the thickness-averaged magnetization ${\langle m_z \rangle}_z$ during the three-step fluctuation-free topological phase transition.}
Trend of the average out-of-plane magnetization ${\langle m_z \rangle}_z$ (black line) and minimum and maximum net magnetization ${\langle m_z \rangle}_z^\text{min/max}$ (blue lines) as a function of the pseudo-temperature $\Theta$ during the adiabatic cooling procedure at $B\SI{=140}{\mT}$ shown in Fig.~\ref{fig:fig3}.
The gray dashed lines mark the critical temperatures for the three transitions: 1. weak stripe formation, 2. bubble nucleation, 3. topological switching.
}
\label{fig:Fig_CC_formation_02-Copy8}
\end{figure*}
 \newpage
 
\begin{figure*}[h]
\includegraphics[width=110 mm]{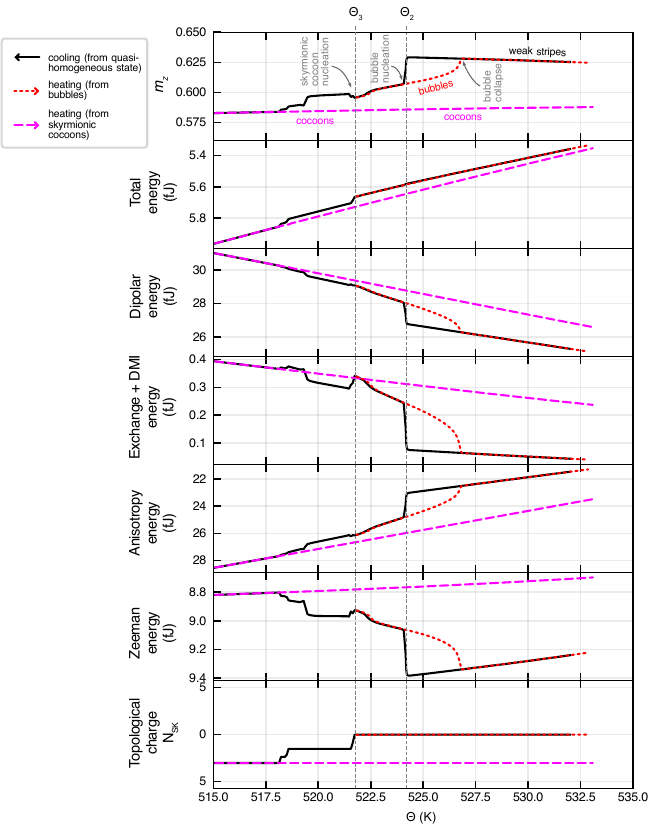}
\caption{
\textbf{Coexistence of weak stripes and skyrmionic cocoon states.}
Trends of the magnetization, micromagnetic energy terms and total topological charge as a function of the pseudo-temperature $\Theta$ during adiabatic cooling.
Note that the only energy term reduced by the transition from weak stripes to cocoons is the interfacial anisotropy energy.
The plots also display the trends for slow heating starting from trivial bubble (dashed line) and skyrmionic cocoon (dotted line) states, which reveal hysteretic behaviour above the critical temperatures $\Theta_\mathrm{2}$ and $\Theta_\mathrm{3}$, in agreement with the first-order nature of the corresponding transitions.
}
\label{fig:Fig_S_energies_coexistence}
\end{figure*}
\newpage

\begin{figure*}[h]
\includegraphics[width=18 cm]{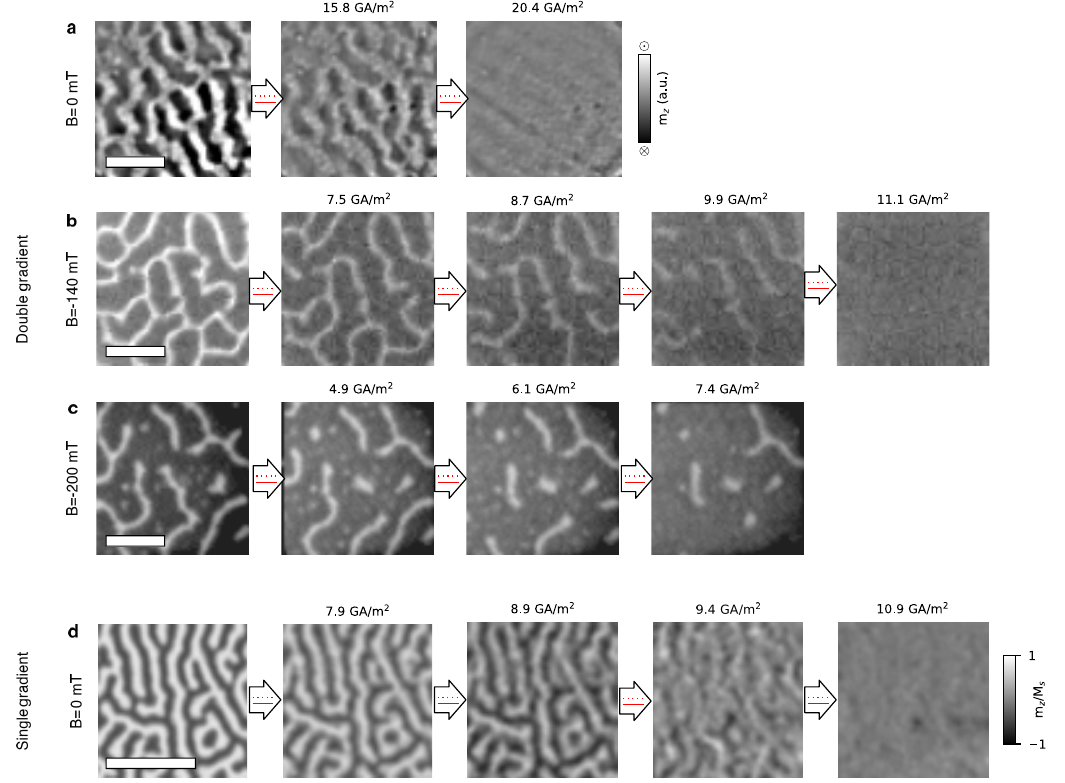}
\caption{
\textbf{High-temperature states in direct-current-heated aperiodic magnetic multilayers.}
X-ray microscopy images of magnetic states obtained by progressively heating aperiodic multilayer samples with increasing direct current densities.
\textbf{a,} DG multilayer ($N_\text{SG}=13$, $d_0=\SI{1.6}{\nm}$, $N_\text{ML}=15$, $d_\text{ML}=\SI{1.5}{\nm}$).
\textbf{b,} DG multilayer ($N_\text{SG}=13$, $d_0=\SI{1.8}{\nm}$, $N_\text{ML}=15$, $d_\text{ML}=\SI{0.9}{\nm}$).
\textbf{c,} DG multilayer ($N_\text{SG}=13$, $d_0=\SI{1.6}{\nm}$, $N_\text{ML}=15$, $d_\text{ML}=\SI{1.0}{\nm}$).
\textbf{d,} SG multilayer ($N_\text{SG}=13$, $d_0=\SI{2.0}{\nm}$).
(\textbf{a}) are FTH reconstructions, (\textbf{b,c}) are STXM images, (\textbf{d}) are x-ray ptychography images.
White scalebars are \SI{1}{\micro\meter}.
}
\label{fig:FigDC_heating}
\end{figure*}
\newpage

\begin{figure*}[h]
\includegraphics[width=12.5 cm]{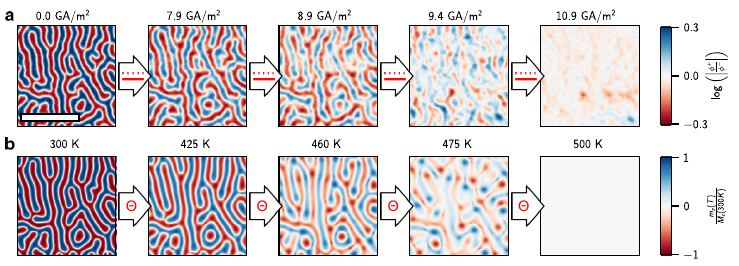}
\caption{
\textbf{Experiment and simulation of magnetic states during adiabatic increase of the temperature}.
\textbf{a,} Experimental data, obtained via x-ray ptychography in a magnetic track heated via direct currents at remanence. Current densities are printed above the images.
Quantitative magnetic contrast was obtained by acquiring images with both left- and right-circularly polarized light and dividing them with each other.
\textbf{b,} Micromagnetic simulations in which the pseudo-temperature $\Theta$ is gradually increased.
Sample structure is $N_\text{SG}=13$, $d_0=\SI{2.0}{\nm}$, $S=\SI{0.1}{\nm}$. White scalebar is \SI{1}{\micro\m}.
}
\label{fig:FIG_SG_heating_vs_T_fromMAXIV}
\end{figure*}
\clearpage
\newpage

\begin{figure*}[h]
\includegraphics[width=12.5 cm]{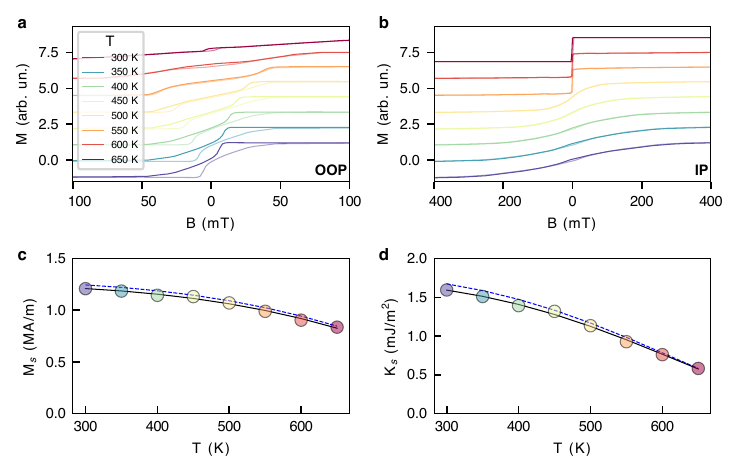}
\caption{
\textbf{SQUID data for a [Pt(\SI{3}{\nm})$\mid$Co(\SI{1.5}{\nm})$\mid$Al(\SI{1.4}{\nm})]\texttimes 3 stack.}
\textbf{a,} Out-of-plane hysteresis loops at increasing temperatures, vertically offset for better visibility.
\textbf{b,} In-plane hysteresis loops at increasing temperatures, also vertically offset.
\textbf{c,} Saturation magnetization $M_\mathrm{s}$ as a function of temperature.
\textbf{d,} Interfacial uniaxial anisotropy $K_\mathrm{s}$ as a function of temperature.
The black line is a fit to the data, while the blue line is a second fit from slightly different values resulting from a different background subtraction process.
Values obtained from the black and blue line fit have been used for the simulations described in Figs.~\ref{fig:fig1},\ref{fig:fig2} and Figs.~\ref{fig:fig3},\ref{fig:fig4}, respectively.
The results and conclusions for both fits are the same.
}
\label{fig:SQUID_vs_T}
\end{figure*}
\clearpage
\newpage

\begin{figure*}[h]
\includegraphics[width=12.5 cm]{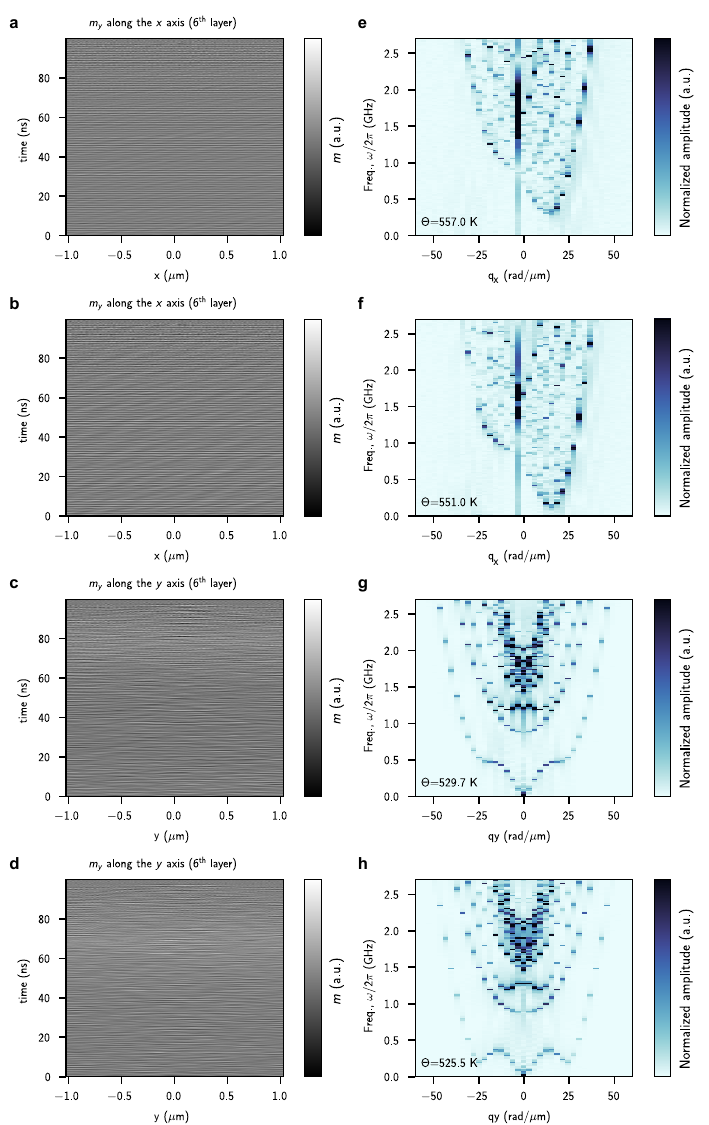}
\caption{\textbf{Examples of simulated magnon band dispersion.}
\textbf{a-d,} In-plane reduced magnetization component $m_y$ on the 6\textsuperscript{th} layer of the single-gradient stack as a function of time along a line parallel to the $x$ (\textbf{a,b}) and $y$ (\textbf{c,d}) axes, for varying pseudo-temperature $\Theta$.
\textbf{a,b} refer to the first transition, \textbf{c,d} to the second.
\textbf{e-h,} 2D fast Fourier transform of the quantities displayed in \textbf{a,d}, resulting in the magnon band dispersion.
The band dispersion of the lowest energy modes and the associated magnon bandgap are displayed in Fig.~\ref{fig:fig3}{i,l}.
}
\label{fig:magnondispersion}
\end{figure*}

\end{document}